\documentclass[journal]{IEEEtran}

\usepackage{cite}
\ifCLASSINFOpdf
  \usepackage[pdftex]{graphicx}
\else
\fi
\usepackage{amsmath}
\usepackage{amssymb}
\usepackage{mathdots}
\usepackage{algorithmic}
\usepackage{algorithm}
\usepackage{array}

\usepackage{cuted}
\usepackage{stfloats}

\ifCLASSOPTIONcompsoc
  \usepackage[caption=false,font=normalsize,labelfont=sf,textfont=sf]{subfig}
\else
  \usepackage[caption=false,font=footnotesize]{subfig}
\fi

\usepackage{tikz}
\usetikzlibrary{external}
\tikzsetexternalprefix{Figures_Generated/}
\usepackage{pgfplots}
\pgfplotsset{compat=newest}
\pgfplotsset{legend style={rounded corners=2pt,nodes=right}}
\DeclareMathOperator{\fieldR}{\mathbb{R}}
\DeclareMathOperator{\fieldC}{\mathbb{C}}

\DeclareMathOperator{\setK}{\mathcal{K}}

\DeclareMathOperator{\T}{\operatorname{T}}
\DeclareMathOperator{\Herm}{\operatorname{H}}
\newcommand{\trans}[1]{#1^{\mathrm{T}}}
\newcommand{\herm}[1]{#1^{\mathrm{H}}}
\newcommand{\ve}[1]{\boldsymbol{#1}}

\newcommand{\e}{\mathrm{e}}
\newcommand{\E}{\mathrm{E}}
\allowdisplaybreaks
\begin{document}
\newlength{\figurewidth}
\newlength{\figureheight}
\title{Joint Transmit and Receive Filter Optimization\\ for Sub-Nyquist Delay-Doppler Estimation}
\author{Andreas~Lenz,~
Manuel~S.~Stein,~
A.~Lee~Swindlehurst%
\thanks{This work was supported by the EIKON e.V., the Heinrich and Lotte M\"{u}hlfenzl Foundation and the Institute for Advanced Study (IAS), Technische Universit\"{a}t M\"{u}nchen (TUM), with funds from the German Excellence Initiative and the European Union's Seventh Framework Program (FP7) under grant agreement no.~291763. This work was also supported by the German Academic Exchange Service (DAAD) with funds from the German Federal Ministry of Education and Research (BMBF) and the People Program (Marie Curie Actions) of the European Union's Seventh Framework Program (FP7) under REA grant agreement no.~605728 (P.R.I.M.E. - Postdoctoral Researchers International Mobility Experience).}
\thanks{A. Lenz is with the Institute for Communications Engineering, Technische Universit\"at M\"unchen, Germany  (e-mail: andreas.lenz@mytum.de). M. S. Stein is with Mathematics Department, Vrije Universiteit Brussel, Belgium and with the Chair for Stochastics, Universit\"at Bayreuth, Germany (e-mails: manuel.stein@vub.ac.be, manuel.stein@uni-bayreuth.de). A. L. Swindlehurst is with the Henry Samueli School of Engineering, University of California, Irvine, USA (e-mail: swindle@uci.edu).}
}

\maketitle

\begin{abstract}
In this article, a framework is presented for the joint optimization of the analog transmit and receive filter with respect to a parameter estimation problem. At the receiver,  conventional signal processing systems restrict the two-sided bandwidth of the analog pre-filter $B$ to the rate of the analog-to-digital converter $f_s$ to comply with the well-known Nyquist-Shannon sampling theorem. In contrast, here we consider a transceiver that by design violates the common paradigm $B\leq f_s$. To this end, at the receiver, we allow for a higher pre-filter bandwidth $B>f_s$ and study the achievable parameter estimation accuracy under a fixed sampling rate when the transmit and receive filter are jointly optimized with respect to the Bayesian Cram\'{e}r-Rao lower bound. For the case of delay-Doppler estimation, we propose to approximate the required Fisher information matrix and solve the transceiver design problem by an alternating optimization algorithm. The presented approach allows us to explore the Pareto-optimal region spanned by transmit and receive filters which are favorable under a weighted mean squared error criterion. We also discuss the computational complexity of the obtained transceiver design by visualizing the resulting ambiguity function. Finally, we verify the performance of the optimized designs by Monte-Carlo simulations of a likelihood-based estimator.
\end{abstract}

\IEEEpeerreviewmaketitle
\section{Introduction}%
\IEEEPARstart{T}{he} inference of unknown parameters is of interest in technical applications such as radar, sonar, image analysis, biomedicine or seismology. In radar systems knowledge of the delay-Doppler shift can be used to determine the distance and velocity of a target object. For imaging and biomedical applications conclusions about the test medium can be drawn from the received signal parameters. During the design of such systems, resource and hardware restrictions, e.g., power consumption and computational effort, have to be taken into consideration and a favorable trade-off between high system performance and low system complexity has to be found.

At the transmitter, using large bandwidths is, in general, favorable for the performance. Concerning the analog hardware costs at the receiver side, in particular, the sampling rate of the analog-to-digital (A/D) conversion has been identified as a bottleneck \cite{Walden99, Verhelst15}. In general, one can take two different engineering perspectives regarding the performance-complexity trade-off associated with the digitization process at the receiver. On the one hand, if the sampling rate $f_s$ is limited, fulfilling the sampling theorem by setting the analog pre-filter bandwidth to $B\leq f_s$ means that the performance is bounded by the complexity of the sampling device. On the other hand, if a certain bandwidth $B$ is available to the transmitter, fulfilling the Nyquist-Shannon theorem requires one to provide sufficiently large power and hardware resources at the receiver to perform sampling with $f_s \geq B$. While choosing $B\leq f_s$ guarantees perfect reconstruction of the band-limited analog received signal from its digital samples, for the signal processing task of pilot-based estimation we are instead interested in the accurate recovery of only a few signal parameters using a probabilistic approach. Since high estimation accuracy rather than a low reconstruction error is the desired goal, compliance with the sampling theorem can be relaxed. This leads to the fundamental question of how to design signal parameter estimation methods and systems when commonly used  principles like the Nyquist-Shannon theorem are set aside.
\section{Problem Formulation}
Before treating the special case of delay-Doppler estimation, we state the considered technical problem generically.
\vspace{-.2cm}
\subsection{Generic System Model}
\begin{figure}[h]
	\centering
	\includegraphics{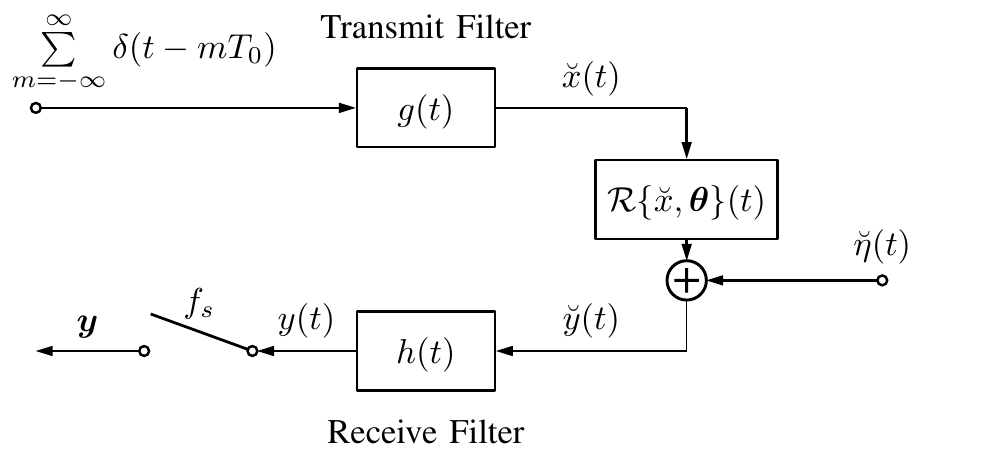}
	\caption{System Model (Baseband Representation)}
	\label{fig:Signal_Flow}
	
\end{figure}

Consider the system model in Fig. \ref{fig:Signal_Flow},
\vspace{-.2cm}
where a periodic analog waveform $\breve{x}(t) \in \mathbb{C}, t \in \mathbb{R}$ is emitted by continuously repeating the waveform $g(t) \in \mathbb{C}$ of bandwidth $B$,
\begin{align}\label{signal:transmitter}
\breve{x}(t) &= \sum_{m=-\infty}^{\infty} g(t-mT_0),
\end{align}
in intervals of $T_0$. The transmit signal \eqref{signal:transmitter} is transformed according to a parametric propagation model $\mathcal{R}\{\breve{x}, \ve{\theta}\}(t)$, which is characterized by a functional operator with parameters $\boldsymbol{\theta} \in \mathbb{R}^{D}$. At the receiver, the waveform is perturbed by additive white Gaussian noise $\breve{\eta}(t) \in \mathbb{C}$ with constant power spectral density $N_0$, resulting in the analog sensor signal $\breve{y}(t)\in\mathbb{C}$,
\begin{equation}\label{signal:receiver:unfiltered}
\breve{y}(t) = \mathcal{R}\{\breve{x}, \ve{\theta}\}(t) + \breve{\eta}(t).
\end{equation}
The signal \eqref{signal:receiver:unfiltered} is filtered by a linear, time-invariant filter $h(t) \in \mathbb{C}$, such that the final analog received signal is given by
\begin{align}
y(t) &= \big(\mathcal{R}\{\breve{x}, \ve{\theta}\}(t) + \breve{\eta}(t) \big)*h(t)\notag\\
&= v(t;\boldsymbol{\theta}) + \eta(t).
\label{eq:Filtered_Receive_Signal}
\end{align}
Note that here the filtering operation with $h(t)$ is performed in the analog domain before the sub-Nyquist A/D conversion. Therefore, it is possible to shape the aliasing effect and diminish the associated information loss. In contrast, a digital receive filter would act on the sampled data and consequently could not recover the loss induced by sampling artifacts.

The signal $y(t)\in\mathbb{C}$ is sampled in intervals of $T_s = \frac{1}{f_s}$ for the duration $T_0$, resulting in $N = \frac{T_0}{T_s} \in \mathbb{N}$ samples
\begin{equation}\label{eq:Sampled_Receive_Signal}
\boldsymbol{y} = \boldsymbol{v} (\boldsymbol{\theta}) + \boldsymbol{\eta},
\end{equation}
where we use vector notation to denote $N$ samples
\begin{align}
\boldsymbol{u} &= \Bigg[u\Bigg(-\frac{N}{2}T_s\Bigg), \dots, u\Bigg(\bigg(\frac{N}{2}-1\bigg)T_s\Bigg)\Bigg]^{\T} \in \mathbb{C}^{N}
\label{eq:Sampled_Signals}
\end{align}
of an analog waveform $u(t)\in\mathbb{C}$ with $t\in[-\frac{T_0}{2}, \frac{T_0}{2})$. We assume that the noise samples $\boldsymbol{\eta} \in \mathbb{C}^{N}$ in \eqref{eq:Sampled_Receive_Signal} follow a multivariate zero-mean Gaussian distribution with covariance
\begin{equation}\label{eq:noise:covariance:matrix}
\boldsymbol{R}_{\boldsymbol{\eta}} = \E_{\boldsymbol{\eta}}[\boldsymbol{\eta}\boldsymbol{\eta}^{\Herm}] \in \mathbb{C}^{N \times N}.
\end{equation}
\subsection{Parameter Estimation Task}
Throughout this work a pilot-based Bayesian estimation approach is assumed, i.e., the receiver has perfect knowledge about the transmit signal $\breve{x}(t)$, the receive filter $h(t)$, the sampling rate $f_s$ and the functional form of $\mathcal{R}\{ \breve{x}, \ve{\theta} \}(t)$ for each $\ve{\theta}$. Therefore, the exact probabilistic model $p(\boldsymbol{y}|\boldsymbol{\theta})$ is available at the receiver, while the propagation parameters $\boldsymbol{\theta}$ are assumed to be unknown random variables distributed according to a known probability law $\boldsymbol{\theta}\sim p(\boldsymbol{\theta})$.

The goal of the receiver is to infer the unknown parameters $\boldsymbol{\theta}$ based on the received digital data $\boldsymbol{y}$ using an appropriate estimation algorithm $\boldsymbol{\hat{\theta}} (\boldsymbol{y})$. The weighted estimation error of the processing procedure $\boldsymbol{\hat{\theta}} (\boldsymbol{y})$ is
\begin{equation}\label{eq:WMSE}
\mathrm{MSE}( \boldsymbol{M} ) = \mathrm{tr} \big( \boldsymbol{M} \boldsymbol{R}_{\ve{\epsilon}} \big), 
\end{equation}
for some positive semidefinite weighting matrix $\boldsymbol{M} \in \mathbb{R}^{D \times D}$, where we define the mean squared error (MSE) matrix by

\begin{equation}\label{eq:MSE:matrix}
\boldsymbol{R}_{\ve{\epsilon}} = \E_{\boldsymbol{y}, \boldsymbol{\theta}} \Big[\big(  \boldsymbol{\hat\theta}(\boldsymbol{y}) - \boldsymbol{\theta} \big) \big( \boldsymbol{\hat\theta}(\boldsymbol{y}) - \boldsymbol{\theta} \big)^{\T}\Big].
\end{equation}

As a direct computation of the MSE matrix \eqref{eq:MSE:matrix} is generally intractable, a performance analysis is usually based on the \textit{Bayesian Cram\'{e}r-Rao lower bound} (BCRLB) \cite[p. 5]{VanTrees07}
\begin{equation}\label{eq:BCRLB}
\boldsymbol{R}_{\ve{\epsilon}} \succeq \ve{J}_B^{-1},
\end{equation}
where $\ve{J}_B$ is the \textit{Bayesian information matrix} (BIM)
\begin{equation}\label{eq:BIM}
\ve{J}_B = \boldsymbol{J}_D + \boldsymbol{J}_P.
\end{equation}
The BCRLB \eqref{eq:BCRLB} forms a fundamental limit for the achievable estimation accuracy \eqref{eq:MSE:matrix}.
The first term on the right-hand side of \eqref{eq:BIM} represents the \textit{expected Fisher information matrix} (EFIM) 
\begin{equation}\label{eq:EFIM}
\boldsymbol{J}_D  = \E_{\boldsymbol{\theta}}\big[\boldsymbol{J}_F(\boldsymbol{\theta})\big],
\end{equation}
with the \textit{Fisher information matrix} (FIM) exhibiting entries
\begin{equation}\label{eq:FIM:entries}
[\boldsymbol{J}_F(\boldsymbol{\theta})]_{ij} = - \E_{\boldsymbol{y}|\boldsymbol{\theta}} \Bigg[ \frac{\partial^2 \ln p (\boldsymbol{y}| \boldsymbol{\theta})}{\partial [\boldsymbol{\theta}]_i \partial [\boldsymbol{\theta}]_j} \Bigg].
\end{equation}
The matrix \eqref{eq:EFIM} characterizes the average information about the parameters $\boldsymbol{\theta}$ that is embodied in the data $\boldsymbol{y}$. For the Gaussian model \eqref{eq:Sampled_Receive_Signal}, the FIM entries \eqref{eq:FIM:entries} are given by \cite[p. 525]{Kay93}
\begin{equation}\label{eq:FI_Entries_General}
[\boldsymbol{J}_F(\boldsymbol{\theta})]_{ij} = 2 \mathrm{Re} \left\{  \herm{\left( \frac{\partial \boldsymbol{v} (\boldsymbol{\theta})}{\partial [\boldsymbol{\theta}]_i} \right)} \boldsymbol{R}_{\boldsymbol{\eta}}^{-1} \left( \frac{\partial \boldsymbol{v} (\boldsymbol{\theta})}{\partial [\boldsymbol{\theta}]_j} \right) \right\}. 
\end{equation}
The second summand in \eqref{eq:BCRLB} is the prior information matrix (PIM) with entries
\begin{equation}
[\boldsymbol{J}_P]_{ij} =  - \E_{\boldsymbol{\theta}}\Bigg[ \frac{\partial^2 \ln p(\boldsymbol{\theta})}{\partial [\boldsymbol{\theta}]_i \partial [\boldsymbol{\theta}]_j} \Bigg].
\end{equation}
It specifies the information about the unknown parameters which is contained in the prior $p(\boldsymbol{\theta})$. 

Accessing and optimizing the system performance \eqref{eq:MSE:matrix} based on a theoretical measure like \eqref{eq:BIM} has the advantage that the achievable accuracy level can be computed analytically without extensive simulations of the algorithm $\boldsymbol{\hat{\theta}} (\boldsymbol{y})$.
\subsection{Transceiver Design Problem}
The design problem of finding analog transmit and receive filters $(g^{\star}(t),h^{\star}(t))$ that minimize the MSE \eqref{eq:WMSE} of the estimation algorithm $\boldsymbol{\hat{\theta}} (\boldsymbol{y})$ under a particular weighting $\ve{M}$, subject to a transmit power constraint $P_T$, can be formulated as
\begin{align}\label{eq:optimization:problem:initial}
\underset{g(t), h(t)}{\arg \min}\,\mathrm{tr} (\boldsymbol{M} \boldsymbol{R}_{\ve{\epsilon}}),\quad\mathrm{s.t.}\quad\frac{1}{T_0} \int_{T_0} |\breve{x}(t)|^2 \mathrm{d}t \leq P_T.
\end{align}
For simplicity, we assume that the system operates in the linear regime of the transmit power amplifier and therefore do not include a peak power constraint.

Although the BRCLB \eqref{eq:BCRLB} can be achieved with equality only under special conditions \cite[p. 5]{VanTrees07}, it closely characterizes the trend of the estimation performance. This is illustrated by simulation of the MSE in Section \ref{sec:simulation:results}. It is therefore possible to formulate \eqref{eq:optimization:problem:initial} based on the BIM \eqref{eq:BIM}
\begin{equation}\label{eq:optimization:problem:BIM}
\underset{g(t), h(t)}{\arg\min}\,\mathrm{tr} (\boldsymbol{M} \ve{J}_B^{-1}),\quad\mathrm{s.t.}\quad\frac{1}{T_0} \int_{T_0} |\breve{x}(t)|^2 \mathrm{d}t \leq P_T.
\end{equation}
To avoid optimization with respect to $\ve{J}_B^{-1}$ in \eqref{eq:optimization:problem:BIM}, we consider an alternative maximization problem
\begin{equation} \label{eq:optimization:problem:max}
\underset{g(t), h(t)}{\arg\max}\,\mathrm{tr} (\boldsymbol{M'} \boldsymbol{J}_{B}),\quad\mathrm{s.t.}\quad\frac{1}{T_0} \int_{T_0} |\breve{x}(t)|^2 \mathrm{d}t \leq P_T,
\end{equation}
which is equivalent to \eqref{eq:optimization:problem:BIM} in a boundary preserving sense. It can be shown that if $(g^\star(t), h^\star(t))$ is a solution of the maximization problem \eqref{eq:optimization:problem:max} with $\ve{M}'$, there exists a weighting matrix $\ve{M}$ (not necessarily equal to $\ve{M}'$) for which the original optimization problem \eqref{eq:optimization:problem:BIM} has the same solution $(g^\star(t), h^\star(t))$ \cite{Stein14b}. Since $\ve{J}_P$ is independent of $g(t)$ and $h(t)$, \eqref{eq:optimization:problem:max} simplifies to
\begin{equation} \label{eq:optimization:problem:efim}
\underset{g(t), h(t)}{\arg\max}\,\mathrm{tr} (\boldsymbol{M'} \boldsymbol{J}_{D}),\quad\mathrm{s.t.}\quad\frac{1}{T_0} \int_{T_0} |\breve{x}(t)|^2 \mathrm{d}t \leq P_T.
\end{equation}
\section{Contribution and Outline}
For the example of a single-input single-output (SISO) system with delay-Doppler shift, we develop a framework to solve the transceiver optimization problem \eqref{eq:optimization:problem:efim} under the sub-Nyquist condition $B>f_s$. Note that such pilot-based delay-Doppler estimation problems are key for wireless applications like radar or radio-based localization and positioning.

After providing an overview on related work (Section \ref{sec:related:work}), we 
\begin{itemize}
\item Introduce the receive model with delay-Doppler shift and derive its FIM under arbitrary transmit and receive filters (Section \ref{section:performance:measure:delay:doppler}).
\item Show how to approximate the information matrix $\boldsymbol{J}_F(\ve{\theta})$ and $\boldsymbol{J}_D$ in the frequency domain (Section \ref{sec:approximation:fim}).
\item Provide an algorithm to solve the transceiver design problem \eqref{eq:optimization:problem:efim} by alternating between optimizing the transmit $g(t)$ and the receive filter $h(t)$ (Section \ref{sec:optimization:algorithm}).
\item Explore the Pareto-optimal region obtained by jointly optimizing the transmit and  receive filter $g(t)$ and $h(t)$ with respect to a weighted MSE criterion (Section \ref{sec:optimization:results}).
\item Discuss the ambiguity function resulting from the optimized transceiver design (Section \ref{sec:optimization:results}).
\item Verify the performance obtained under an optimized design by Monte-Carlo simulation of an asymptotically efficient algorithm $\boldsymbol{\hat{\theta}} (\boldsymbol{y})$ (Section \ref{sec:simulation:results}).
\end{itemize}
Our final conclusions are outlined in Section \ref{sec:conclusion}. Note that preliminary results have been discussed in \cite{Stein14a, Lenz15, Lenz17}, where we have focused on either optimizing the receive or the transmit filter of a sub-Nyquist system, and \cite{Khayambashi14} which was centered around a compressive sensing framework.

\section{Related Work}%
\label{sec:related:work}
The analysis and optimization of wireless transceiver systems enjoys significant attention in the signal processing community \cite{Yang07,Haykin08, Romero11, Stoica12}.
While for example \cite{Bell93,Guerci00} focus on signal optimization for classical radar systems, waveform design of orthogonal frequency division multiplexing (OFDM) signals for radar applications have been considered in \cite{Calderbank09, Berger10}. The favorable design of satellite-based positioning signals with high time-delay estimation accuracy is discussed in \cite{Antreich11}.
Optimization of transmit and receive filters in the spatial domain plays a crucial role in multiple-input multiple-output (MIMO) radar systems \cite{Fishler04, Stoica07}. For cognitive radars, high accuracy is achieved by using a Bayesian tracking algorithm and adapting the transmit signal to the state-space knowledge \cite{Haykin06}. In \cite{Huleihel13} the concept of cognitive radar is extended to MIMO radar systems resulting in improved parameter identifiability and higher resolution \cite{Bliss03} in comparison to phased arrays. Further, MIMO radar features the effect of virtual receive antennas \cite{Li07}, providing high accuracy at lower hardware costs.

When performing the transceiver system optimization various figures of merit have been considered. In particular mutual information and signal-to-noise ratio (SNR) are popular metrics \cite{Romero11,Bell93, Guerci00,Leshem07}, while a low MSE for the estimated parameters is an alternative choice and leads to the use of estimation theoretic error bounds \cite{Li08,Hurtado08,Huleihel13}. Further, the ambiguity function (AF) of a given transmit signal is a classical tool to characterize the waveform quality for applications where joint estimation of time and frequency shifts is desired \cite{Wilcox60, Knapp76,Stein81,Fishler04,Antonio07,Chen08}.
Subspace-based techniques for delay-Doppler estimation problems are presented in \cite{Swindlehurst98b,Jakobsson98} and lead to significant complexity reduction of the maximum-likelihood (ML) estimator. 

An important aspect in the context of this article is the field of compressed sensing (CS) and finite rate of innovation (FRI) techniques, where sparse signal structures \cite{Donoho06} and signals with finite degrees of freedom \cite{Vetterli02} are exploited to facilitate signal processing. CS-related work focuses on methods for reconstructing signals that embody sparsity from a small number of samples \cite{Candes08}, while FRI techniques attempt to reconstruct the original signal without bandwidth limitations \cite{Wei14}. In \cite{Maravic05,Tan08,Ben-Haim12} reconstruction of FRI signals in noise is considered. 
The work of  \cite{Davenport10} describes CS methods for parameter estimation applications where perfect signal reconstruction is not required.
Taking into account sampling hardware complexity, Xampling has been introduced as a sub-Nyquist sampling technique on multiple analog projections of the received signal \cite{Mishali11a,Michaeli12}. With such an approach, low-rate analog-to-digital conversion architectures for radar systems have been proposed which exhibit the same estimation performance as receivers that operate at the Nyquist rate when the SNR is sufficiently high \cite{Mishali11b,Baransky14}. In \cite{Gu17} an information-theoretic method is presented to optimize the analog projection functions for compressed sub-Nyquist sampling within a Bayesian estimation framework. It is shown that the MSE of the resulting time-delay estimate can be improved for the undersampled system compared with Nyquist sampling methods.
In contrast to these works, we consider likelihood-oriented signal processing and transceiver optimization while using a conventional baseband receiver which operates at a sampling rate $f_s$ smaller than the Nyquist bandwidth $B$.
\section{Delay-Doppler Estimation}
\label{section:performance:measure:delay:doppler}
\subsection{Specific Receive Model}
To illustrate a specific case of the generic transceiver design problem \eqref{eq:optimization:problem:efim}, we consider a single-path propagation model,
\begin{equation}\label{eq:Delay_Doppler_Definition}
\mathcal{R}\{\breve{x}, \ve{\theta}\}(t) = \gamma \breve{x}(t-\tau) \e^{\mathrm{j} 2\pi \nu t},
\end{equation}
with path gain $\gamma \in \mathbb{C}$, time-delay $\tau \in \mathbb{R}$ and Doppler shift $\nu \in \mathbb{R}$. Such a receive model is often employed in radar systems, where the signal is delayed according to the distance of the reflector and frequency shifted (Doppler effect) due to its relative velocity. For solving the filter design problem \eqref{eq:optimization:problem:efim}, we ignore $\gamma$ and treat it as known and deterministic, so that the unknown random parameter vector is
\begin{equation}
\boldsymbol{\theta} = \begin{bmatrix}\tau &\nu\end{bmatrix}^\mathrm{T}.
\end{equation}
Several reasons motivate this approach. First, accurate estimation of the linear coefficient $\gamma$ is less crucial for most applications where the delay and Doppler are used for ranging and localization, and thus we focus on the optimization of the estimation accuracy for these parameters.  This leads to a lower BCRLB for $\tau$ and $\nu$ than for the case where $\gamma$ is treated as unknown [3, p. 48]. The resulting BCRLB still serves as a valid lower bound for the estimation error of $\tau$ and $\nu$. Second, considering $\gamma$ as known simplifies the optimization of the BCRLB and reduces the computational effort when evaluating the corresponding Pareto-optimal region.  In practice, $\gamma$ is a nuisance parameter that must still be estimated.  We will illustrate in Section \ref{sec:simulation:results} that the MSE for $\tau$ and $\nu$ is still very close to the smaller BCRLB even when $\gamma$ must be estimated. This indicates that there is little loss when ignoring $\gamma$ during the optimization of the transceive filters $g(t)$ and $h(t)$.

In the following it is assumed that the unknown parameters are Gaussian distributed
\begin{equation}
\boldsymbol{\theta} \sim \mathcal{N}(\boldsymbol{0}, \ve{R}_{\ve{\theta}})
\end{equation}
and stochastically independent, such that the covariance matrix of the prior is
\begin{equation}\label{covariance:matrix:prior}
\ve{R}_{\ve{\theta}} = \begin{bmatrix}
\sigma^2_\tau & 0 \\
0 & \sigma^2_\nu
\end{bmatrix}.
\end{equation}
\subsection{Frequency Domain Representation}
Solving the optimization problem \eqref{eq:optimization:problem:efim} requires an analytical characterization of the EFIM \eqref{eq:EFIM} under the receive model \eqref{eq:Delay_Doppler_Definition}. A frequency-domain representation enables a compact and insightful notation for the received signal model \cite{Lenz15,Zeira90} and thus of the FIM entries \eqref{eq:FI_Entries_General}. 

For the derivation, we assume a fixed sampling rate $f_s$ at the receiver. The transmit filter $g(t)$ is band-limited with two-sided bandwidth $B$. In contrast to the common assumption $B \leq f_s$, in our setup we allow $B > f_s$. Due to periodicity, the waveform $\breve{x}(t)$ can be represented by its Fourier series
\begin{equation}
\breve{x}(t) = \sum_{k=-\frac{K}{2}}^{\frac{K}{2}-1} G_k \e^{\mathrm{j} k \omega_0 t}, \label{eq:Periodic_Transmit_Signal}
\end{equation}
where $\omega_0=\frac{2\pi}{T_0} = 2\pi f_0$. Note that $K= \lceil \frac{2\pi B}{\omega_0} \rceil \in \mathbb{N}$ is the total number of harmonics. $G_k$ denotes the $k$-th Fourier coefficient of the transmit filter
\begin{equation}
G_k = \frac{1}{T_0} \int\limits_{-\infty}^{\infty} g(t) \e^{\mathrm{j} k \omega_0 t} \mathrm{d}t.
\end{equation}
We write $\setK$ for the set of Fourier coefficient indices
\begin{equation}
\setK = \Big\{-\frac{K}{2}, -\frac{K}{2}+1, \dots, \frac{K}{2}-1 \Big\}
\end{equation}
with cardinality $|\setK| = K$. Inserting expression (\ref{eq:Periodic_Transmit_Signal}) into (\ref{eq:Delay_Doppler_Definition}) and applying the filtering operation in (\ref{eq:Filtered_Receive_Signal}), we obtain
\begin{align}\label{transmit:signal:receiver:filtered:exact:fdomain}
v(t;\boldsymbol{\theta}) &= \gamma \sum_{k\in \setK} G_k \Big(\e^{\mathrm{j}k\omega_0(t-\tau)} \e^{\mathrm{j} 2 \pi \nu t}\Big) * h(t) \nonumber \\
&= \gamma \e^{\mathrm{j}2\pi \nu t} \sum_{k\in \setK} \e^{\mathrm{j} k\omega_0 t} \e^{-\mathrm{j}k\omega_0 \tau} G_k H(k\omega_0 + 2\pi \nu),
\end{align}
where
\begin{align}
H(\omega)=  \int\limits_{-\infty}^{\infty} h(t) \e^{-\mathrm{j} \omega t} \mathrm{d}t
\end{align}
is the Fourier transform of the receive filter $h(t)$.
Evaluating $v(t;\boldsymbol{\theta})$ at sampling instants $nT_s, n \in \{-\frac{N}{2}, \dots, \frac{N}{2}-1\}$ yields
\begin{equation}\label{eq:transmit:signal:sampled:no_alias}
v(nT_s;\boldsymbol{\theta}) = \gamma \e^{\mathrm{j} 2\pi \nu n T_s} \sum_{k\in \setK} \e^{\mathrm{j}2\pi\frac{kn}{N}} \e^{-\mathrm{j}k\omega_0 \tau} G_k H(k\omega_0 + 2\pi \nu).
\end{equation}
The set $\setK$ can be partitioned into $N$ disjoint subsets $\mathcal{F}_k, k \in \{-\frac{N}{2}, \dots, \frac{N}{2}-1\}$, where each $\mathcal{F}_k$ contains elements of $\setK$ in intervals of $N$ and $k \in \mathcal{F}_k$. Denote by $L^-_k$ the integer such that $k-NL^-_k$ is the smallest element of $\mathcal{F}_k$, and respectively $L^+_k$ the integer such that $k+NL^+_k$ is the largest element of the subset $\mathcal{F}_k$. The subsets $\mathcal{F}_k$ then can be denoted as
\begin{align}
\mathcal{F}_k = \left\{ k-NL^-_k, \dots,  k, k + N, \dots, k+NL^+_k \right\}.
\end{align}
Splitting the sum over $k$ in (\ref{eq:transmit:signal:sampled:no_alias}) according to this partition and further using $\omega_s = 2\pi f_s = N\omega_0$ yields
\begin{align}\label{eq:transmit:signal:sampled:fdomain}
v(nT_s;\boldsymbol{\theta})=& \gamma \e^{\mathrm{j}2\pi \nu nT_s} \sum_{k=-\frac{N}{2}}^{\frac{N}{2}-1} \e^{\mathrm{j}2\pi \frac{kn}{N}} \sum_{l= -L^-_k}^{L^+_k} \e^{-\mathrm{j}(k\omega_0 + l\omega_s)\tau} \nonumber \cdot \\
& \qquad \qquad G_{k-lN} H(k\omega_0 +l\omega_s + 2\pi \nu).
\end{align}
Here $L^-_k$ and $L^+_k$ can be interpreted as the number of signal spectrum points in the negative, respectively positive frequency domain that superpose at the frequency point $k\omega_0$ due to sampling at a rate $\omega_s$. These limits are given by
\begin{align}
L^-_k &= \left\lfloor \frac{\frac{K}{2} + k}{N} \right\rfloor, \label{eq:L_n} \\
L^+_k &= \left\lfloor \frac{\frac{K}{2} -1 - k}{N} \right\rfloor. \label{eq:L_p}
\end{align}

Note that if the bandwidth is an odd multiple of the sampling frequency $f_s$, i.e,
\begin{equation}
B= 2Lf_s +Nf_0 = \left(2L + 1\right)f_s , \quad L \in \mathbb{N}_0,
\end{equation}
the number of overlapping frequencies in the positive and negative frequency domain adjoin and do not depend on the center $k$. In this case $L^-_k$ and $L^+_k$ are
\begin{equation}
L^-_k = L^+_k = L.
\end{equation}
Stacking the entries \eqref{eq:transmit:signal:sampled:fdomain} into a vector yields
\begin{equation}\label{eq:v_No_Approx}
\boldsymbol{v}(\boldsymbol{\theta}) = \gamma\sqrt{N}\boldsymbol{\Delta}(\nu) \boldsymbol{W}^{\mathrm{H}} \boldsymbol{A} \boldsymbol{T}(\tau) \big(\boldsymbol{\tilde{g}} \circ \boldsymbol{\tilde{h}}(\nu) \big). 
\end{equation}
Denoting $\circ$ as the element-wise Hadamard product, (\ref{eq:v_No_Approx}) contains an aliasing matrix $\boldsymbol{A} \in \{0,1\}^{N \times K}$ with entries
\begin{align}\label{eq:def_A_matrix}
[\boldsymbol{A}]_{ij} = \left\{ \begin{array}{ll}1, & \left(j-\frac{K}{2}-1\right) \in 
\mathcal{F}_{i-\frac{N}{2}-1} \\0, & \mathrm{otherwise}
\end{array} \right.,
\end{align}
a diagonal time-delay matrix $\boldsymbol{T}(\tau) \in \mathbb{C}^{K \times K}$ with components
\begin{align}\label{eq:tau:matrix:entries}
[\boldsymbol{T}(\tau)]_{ii} &= \e^{-\mathrm{j} \left(i-\frac{K}{2}-1\right) \omega_0 \tau},
\end{align}
and the transmit filter spectrum vector $\boldsymbol{\tilde{g}}\in \mathbb{C}^{K}$ with elements
\begin{align}\label{frequency:representation:transmit:filter}
[\boldsymbol{\tilde{g}}]_i &= G_{i-\frac{K}{2}-1}.
\end{align}
Throughout this work vector and matrix entries are referenced by positive integers, starting at $1$ and therefore $i\in \{1, 2, \dots, N\}, j\in \{1, 2, \dots, K\}$ in equation (\ref{eq:def_A_matrix}) and $i\in \{1, 2, \dots, K\}$ in both preceding equations. The vector $\boldsymbol{\tilde{h}}(\nu)\in \mathbb{C}^{K}$ in \eqref{eq:v_No_Approx} denotes the frequency shifted receive filter spectrum with entries
\begin{align}\label{eq:h_d_tilde}
[\boldsymbol{\tilde{h}}(\nu)]_i &= H\Bigg( \bigg(i-\frac{K}{2}-1\bigg)\omega_0 + 2\pi \nu \Bigg).
\end{align}
Further, in \eqref{eq:v_No_Approx} we write $\boldsymbol{W} \in \mathbb{C}^{N \times N}$ for the DFT matrix with entries
\begin{align}
[\boldsymbol{W}]_{ij} &= \frac{1}{\sqrt{N}} \e^{-\mathrm{j}2\pi \frac{\left( i-\frac{N}{2} -1 \right) \left(j-\frac{N}{2}-1 \right) }{N}},
\end{align}
where $i,j \in \{1, 2, \dots, N\}$ and $\boldsymbol{\Delta}(\nu) \in \mathbb{C}^{N \times N}$ represents the diagonal matrix with elements
\begin{align}
[\boldsymbol{\Delta}(\nu)]_{ii} &= \e^{\mathrm{j}2\pi \left(i-\frac{N}{2}-1\right) \nu T_s}.
\end{align}

Introducing the Doppler convolution matrix 
\begin{equation}
\boldsymbol{\tilde{\Delta}}(\nu) = \boldsymbol{W} \boldsymbol{\Delta}(\nu) \herm{\boldsymbol{W}} \in \fieldC^{N \times N}, \label{eq:Delta_Tilde}
\end{equation}
a signal representation in the frequency domain 
\begin{align}\label{eq:transmit:signal:sampled:filter:fdomain}
\boldsymbol{v}(\boldsymbol{\theta}) 
&=\gamma\sqrt{N}\herm{\boldsymbol{W}}\boldsymbol{\tilde{\Delta}}(\nu) \boldsymbol{A} \boldsymbol{T}(\tau) \big(  \boldsymbol{\tilde{g}} \circ \boldsymbol{\tilde{h}}(\nu)   \big)\notag\\
&=\sqrt{N} \herm{\boldsymbol{W}} \boldsymbol{\tilde{v}}(\boldsymbol{\theta})
\end{align}
is obtained from \eqref{eq:v_No_Approx}, where
\begin{equation}
\boldsymbol{\tilde{v}}(\boldsymbol{\theta}) = \gamma \boldsymbol{\tilde{\Delta}}(\nu) \boldsymbol{A} \boldsymbol{T}(\tau) \big(  \boldsymbol{\tilde{g}} \circ \boldsymbol{\tilde{h}}(\nu)  \big)
\end{equation}
is the attenuated and delay-Doppler shifted received signal spectrum.
A frequency domain representation of the noise covariance matrix \eqref{eq:noise:covariance:matrix} is obtained by the transformation
\begin{equation}\label{eq:noise:covariance:fdomain}
\boldsymbol{\tilde{R}}_{\boldsymbol{\eta}} = \boldsymbol{W} \boldsymbol{R}_{\boldsymbol{\eta}} \herm{\boldsymbol{W}}.
\end{equation}
Note that due to the Doppler shift in \eqref{eq:h_d_tilde}, frequencies $k\omega_0 + 2\pi \nu$ of the receive filter $H(\omega)$ have to be considered for evaluation of the filtered and sampled signal $\boldsymbol{v}(\boldsymbol{\theta})$. The entries of the FIM $\ve{J}_F(\ve{\theta})$ are directly obtained by plugging \eqref{eq:transmit:signal:sampled:filter:fdomain} into \eqref{eq:FI_Entries_General} and can be found in Appendix \ref{sec:Fisher_Information_Delay-Doppler_Channel}.
\section{Approximation of the Delay-Doppler FIM}
\label{sec:approximation:fim}
These expressions for the FIM make it possible to formulate the transceiver optimization problem \eqref{eq:optimization:problem:efim} for the delay-Doppler model \eqref{eq:Delay_Doppler_Definition}. Due to the assumption of a periodic transmit sequence $\breve{x}(t)$, the minimization over the transmit filter $g(t)$ reduces to an optimization with respect to the $K$ entries of $\ve{\tilde{g}}$. However, using the signal characterization \eqref{eq:transmit:signal:sampled:filter:fdomain} does not lead to an analytically tractable optimization problem, as it is a function of the filter spectrum $H(\omega)$ evaluated at frequencies $k\omega_0 + 2\pi \nu$ (see, e.g., \eqref{eq:transmit:signal:sampled:no_alias}). By the definition of the EFIM \eqref{eq:EFIM} this leads to an objective function which is the expected value of a function that depends on $H(k\omega+2\pi\nu)$ with continuous argument $\nu$. Therefore in the following section we discuss several steps to approximate the FIM entries and formulate a tractable optimization problem \eqref{eq:optimization:problem:efim}. The aim of this section is to simplify the optimization problem so that it can be solved analytically. We then illustrate in Section \ref{sec:optimization:results} that the optimization over the approximated EFIM leads to a significant gain for the exact EFIM. We further show that the region spanned by the approximated EFIM and the exact EFIM exhibit close correspondence. In Section \ref{sec:simulation:results}, we finally show that for the considered delay-Doppler estimation problem the BCRLB, which by definition involves the exact EFIM, accurately characterizes the performance of the MAP estimator.
\subsection{FIM Approximation - Periodic Doppler}
To decouple the receive filter variable $\ve{\tilde{h}}(\nu)$ and the propagation parameter $\nu$, we approximate the Doppler shift to be periodic with $T_0$, i.e.,
\begin{equation}\label{eq:Doppler_Approximation}
\mathrm{e}^{\mathrm{j} 2\pi \nu t} \approx \mathrm{e}^{\mathrm{j}2\pi \nu (t \, \mathrm{modc} \, T_0) }, 
\end{equation}
where we use a centered version of the modulo operation
\begin{equation}
t \, \mathrm{modc} \, T_0 = \left(\left(t + \frac{T_0}{2}\right) \, \mathrm{mod} \, T_0\right) -\frac{T_0}{2}.
\end{equation}
Note that in the interval $t\in\big[-\frac{T_0}{2}, \frac{T_0}{2}\big)$ the approximation \eqref{eq:Doppler_Approximation} is exact. However, due to the filtering operation $h(t)$ portions of the signal outside that time range are convolved into the sampling interval and thus produce an approximation error. 

Under the assumption \eqref{eq:Doppler_Approximation}, it is possible to write
\begin{equation}
\mathrm{e}^{\mathrm{j}2\pi \nu (t \, \mathrm{modc} \, T_0) } = \sum_{z=-\infty}^{\infty} d_z(\nu) \mathrm{e}^{\mathrm{j}z\omega_0t}.
\end{equation}
Calculating the Fourier coefficients $d_z(\nu)$ yields
\begin{equation}
d_z(\nu) =  \left\{ \begin{array}{ll}
	\frac{\sin(\pi T_0 \nu - \pi z)}{\pi T_0 \nu - \pi z}, & z \neq T_0 \nu \\
	1, & z = T_0 \nu
	\end{array} \right. .
\end{equation}
Replacing the Doppler shift in \eqref{eq:Delay_Doppler_Definition} with \eqref{eq:Doppler_Approximation} yields the signal approximation
\begin{align}
&\bar{v}(t;\ve{\theta}) = \gamma \sum_{k\in\setK} \sum_{z=-\infty}^{\infty} G_k d_z(\nu) \mathrm{e}^{-\mathrm{j} k\omega_0 \tau} \mathrm{e}^{\mathrm{j}(k+z)\omega_0t} * h(t) \nonumber \\
&= \gamma \sum_{k\in\setK} \sum_{z=-\infty}^{\infty} G_k d_z(\nu) \mathrm{e}^{-\mathrm{j} k\omega_0 \tau} \mathrm{e}^{\mathrm{j}(k+z)\omega_0t} H\big((k+z)\omega_0\big) \nonumber \\
&= \gamma \sum_{k\in\setK} \sum_{m=-\infty}^{\infty} G_k d_{m-k}(\nu) \mathrm{e}^{-\mathrm{j} k\omega_0 \tau} \mathrm{e}^{\mathrm{j}m\omega_0t} H(m\omega_0),
\end{align}
with the substitution $z = m-k$. Using the abbreviation
\begin{equation}
H_m = H(m\omega_0),
\end{equation}
the signal samples are given by
\begin{align}
\bar{v}(nT_s;\ve{\theta}) = \gamma& \sum_{m=-\infty}^{\infty} \mathrm{e}^{\mathrm{j}2\pi  \frac{mn}{N}} H_m \sum_{k\in\setK} G_k d_{m-k}(\nu) \mathrm{e}^{-\mathrm{j} k\omega_0 \tau} \nonumber \\
= \gamma &\sum_{m = -\frac{N}{2}}^{\frac{N}{2}-1} \mathrm{e}^{\mathrm{j}2\pi  \frac{mn}{N}} \sum_{l=-\infty}^{\infty} H_{m+lN}\cdot \nonumber \\
&\sum_{k\in\setK} G_k d_{m+lN-k}(\nu) \mathrm{e}^{-\mathrm{j} k\omega_0 \tau}.
 \end{align}

Neither the time-delay nor the Doppler shift have a significant impact on the signal bandwidth $B$. The time-delay merely adds a linear phase to the signal spectrum and the Doppler stretch shifts the spectrum by $2 \pi \nu$ (see, e.g., \eqref{eq:transmit:signal:sampled:no_alias}). Since $2\pi \nu \ll \omega_0$ for practical applications, the receive filter can be assumed to have bandwidth $B$. It follows that $H_k = 0$ for all $k$ with $|k\omega_0| > \pi B$. Hence,
\begin{align}
\bar{v}(nT_s;\ve{\theta}) =  \gamma & \sum_{m=-\frac{N}{2}}^{\frac{N}{2}-1} \mathrm{e}^{\mathrm{j}2\pi  \frac{mn}{N}} 
\sum_{l=-L^-_k}^{L^+_k} H_{m+lN} \cdot \nonumber \\ & \sum_{k\in\setK} G_k d_{m+lN-k}(\nu) \mathrm{e}^{-\mathrm{j} k\omega_0 \tau}.
\end{align}
In vector notation, we obtain
\begin{align}
\ve{\bar{v}}(\ve{\theta}) &= \sqrt{N} \gamma \ve{W}^{\mathrm{H}} \ve{A} \big(  \ve{\tilde{h}} \circ \ve{D}(\nu) \ve{T}(\tau) \ve{\tilde{g}} \big)\notag\\
&= \sqrt{N} \herm{\ve{W}} \ve{\tilde{\bar{v}}}(\ve{\theta}), \label{eq:Doppler_Approximation_Signal_Expression}
\end{align}
where the approximated received signal spectrum is
\begin{align}
\ve{\tilde{\bar{v}}}(\ve{\theta}) &= \gamma \ve{A} (  \ve{\tilde{h}} \circ \ve{D}(\nu) \ve{T}(\tau) \ve{\tilde{g}} ),
\end{align}
with the Doppler convolution matrix 
\begin{align}
\ve{D}(\nu) &= \left( \begin{array}{cccc}
d_0(\nu) & d_{-1}(\nu) & \dots & d_{-K+1}(\nu) \\
d_1(\nu) & d_{0}(\nu) & \dots & d_{-K+2}(\nu) \\
\vdots & \vdots & \ddots & \vdots \\
d_{K-1}(\nu) & d_{K-2}(\nu) & \dots & d_{0}(\nu) \\
\end{array} \right)\in \mathbb{C}^{K \times K}
\end{align}
and the receive filter spectrum $\ve{\tilde{h}} \in \fieldC^K$
\begin{align}\label{frequency:representation:receive:filter:simplified}
[\ve{\tilde{h}}]_i &= H\bigg( \Big(i-\frac{K}{2}-1\Big)\omega_0 \bigg).
\end{align}
The advantage of the periodic Doppler approximation \eqref{eq:Doppler_Approximation} is that the filter spectrum $\ve{\tilde{h}}$ does not depend on the Doppler $\nu$. This implies that due to the periodicity of $\bar{v}(t;\ve{\theta})$ only the harmonics $k\omega_0$ of the receive filter $H(\omega)$ are excited and other parts of the filter spectrum do not impact the description of the received signal. Defining the matrix
\begin{equation}
\ve{C}(\ve{\theta}) = \gamma\ve{D}(\nu) \ve{T}(\tau), 
\end{equation}
the received signal spectrum can be represented by
\begin{equation}
\ve{\tilde{\bar{v}}}(\ve{\theta}) = \ve{A} \big(\ve{\tilde{h}} \circ \ve{C}(\ve{\theta}) \ve{\tilde{g}} \big).
\end{equation}
\subsection{FIM Approximation - Circular Noise Covariance}
Computation of the FIM entries \eqref{eq:FI_Entries_General} requires the inverse of the noise covariance matrix $\ve{R}_{\ve{\eta}}^{-1}$. The fact that $\ve{R}_{\ve{\eta}}$ is a non-trivial function of the receive filter $h(t)$ hinders tackling the optimization problem \eqref{eq:optimization:problem:efim}. We exploit the fact that the noise covariance $\ve{R}_{\ve{\eta}}$ is approximately diagonalized by the DFT matrix for large $N$ \cite{Zeira90}, i.e.,
\begin{equation}
\ve{\Omega}_{\ve{\eta}} \approx \ve{W} \ve{R}_{\ve{\eta}}\ve{W}^{\mathrm{H}}, \label{eq:noise_diagonalization}
\end{equation}
where $\ve{\Omega}_{\ve{\eta}} \in \fieldR^{N \times N}$ is a diagonal matrix with entries
\begin{align}
[\ve{\Omega}_{\ve{\eta}}]_{ii} = \frac{1}{T_s} \sum_{l=-L^-_k}^{L^+_k} \left|H\left(\left(i-\frac{N}{2}-1\right)\omega_0 +l\omega_s\right)\right|^2.
\end{align}
Note that \eqref{eq:noise_diagonalization} results from Szeg\"os theorem \cite{Szego58} applied to Toeplitz matrices. Defining the receive filter spectrum matrix
\begin{equation}
\ve{H} = \operatorname{diag}(\ve{\tilde{h}}) \in \fieldC^{K \times K},
\end{equation}
we directly find the useful expression
\begin{equation}
\ve{\Omega}_{\ve{\eta}} = \frac{1}{T_s}  \ve{AH}\herm{\ve{H}} \herm{\ve{A}}.
\end{equation}
\subsection{Approximated FIM Entries}
With the approximations \eqref{eq:Doppler_Approximation} and \eqref{eq:noise_diagonalization}, the FIM elements \eqref{eq:FIM:entries} are directly obtained by inserting \eqref{eq:Doppler_Approximation_Signal_Expression} and \eqref{eq:noise_diagonalization} into \eqref{eq:FI_Entries_General}
\begin{align}
&[\ve{\bar{J}}_F(\ve{\theta})]_{ij} = 2\mathrm{Re} \Bigg\{N \frac{\partial\herm{\ve{\tilde{\bar{v}}}}(\ve{\theta})}{\partial [\ve{\theta}]_i} \ve{\Omega}_{\ve{\eta}}^{-1} \frac{\partial\ve{\tilde{\bar{v}}}(\ve{\theta})}{\partial [\ve{\theta}]_j} \Bigg\} = \nonumber \\ &= 2N\mathrm{Re} \Bigg\{ \herm{\bigg(\ve{\tilde{h}} \circ  \frac{\partial \ve{C}(\ve{\theta}) }{\partial [\ve{\theta}]_i}  \ve{\tilde{g}} \bigg)} \herm{\ve{A}} \ve{\Omega}_{\ve{\eta}}^{-1}  \ve{A} \bigg(\ve{\tilde{h}} \circ  \frac{\partial  \ve{C}(\ve{\theta}) }{\partial [\ve{\theta}]_j} \ve{\tilde{g}} \bigg) \Bigg\}, \label{eq:Doppler_Approximation_Fim_Entries}
\end{align}
where $\ve{\bar{J}}_F(\ve{\theta})$ denotes the FIM that stems from the periodic Doppler and noise covariance approximation. Using the diagonalization property for the Hadamard product of two vectors, equation \eqref{eq:Doppler_Approximation_Fim_Entries} can be denoted by
\begin{equation}\label{eq:FIM:compact:transmit:spectrum}
[\ve{\bar{J}}_F(\ve{\theta})]_{ij} = \ve{\tilde{g}} ^\mathrm{H} \big(\ve{\Phi}_{ij}(\ve{\theta}) + \ve{\Phi}_{ji}(\ve{\theta})\big)  \ve{\tilde{g}} ,
\end{equation}
with
\begin{align}\label{eq:matrix:gamma:ij}
\ve{\Phi}_{ij}(\ve{\theta}) = N  \frac{\partial \herm{\ve{C}}(\ve{\theta}) }{\partial [\ve{\theta}]_i} \herm{\ve{H}} \herm{\ve{A}} \ve{\Omega}_{\ve{\eta}}^{-1} \ve{A} \ve{H} \frac{\partial \ve{C}(\ve{\theta}) }{\partial [\ve{\theta}]_j} \in \fieldC^{K \times K}.
\end{align}
The FIM expression \eqref{eq:FIM:compact:transmit:spectrum} allows us to approximate the objective function \eqref{eq:optimization:problem:efim} by
\begin{align}\label{eq:objective:function:quadratic:transmitter}
\mathrm{tr}(\ve{M}'\ve{J}_D) \approx \mathrm{tr}(\ve{M}'\ve{\bar{J}}_D)
&= \sum_{i=1}^{2} \sum_{j=1}^{2} [\ve{M}']_{ji}[\ve{\bar{J}}_D]_{ij} \notag\\
&= \sum_{i=1}^{2} \sum_{j=1}^{2} [\ve{M}']_{ji}\E_{\ve{\theta}}\left\{ [\ve{\bar{J}}_F(\ve{\theta})]_{ij}  \right\}\notag\\
&=  \ve{\tilde{g}} ^\mathrm{H} \ve{\Phi} \ve{\tilde{g}} ,
\end{align}
with
\begin{equation}\label{eq:matrix:gamma}
\ve{\Phi} = \sum_{i=1}^{2} \sum_{j=1}^{2} [\ve{M}']_{ji}\E_{\ve{\theta}}\left[ \ve{\Phi}_{ij}(\ve{\theta}) + \ve{\Phi}_{ji}(\ve{\theta}) \right].
\end{equation}
Alternatively to \eqref{eq:FIM:compact:transmit:spectrum}, it is possible to emphasize the dependence of the FIM on $\ve{\tilde{h}}$ and reformulate \eqref{eq:Doppler_Approximation_Fim_Entries} as
\begin{equation}
[\ve{\bar{J}}_F(\ve{\theta})]_{ij} = 2 \mathrm{Re} \left\{ \herm{\big(\ve{A} (\ve{\xi}_i(\ve{\theta}) \circ \ve{\tilde{h}})\big)} \ve{\Omega}_{\ve{\eta}}^{-1} \ve{A} (\ve{\xi}_j(\ve{\theta}) \circ \ve{\tilde{h}}) \right\},
\end{equation}
where
\begin{equation}
\ve{\xi}_i(\ve{\theta}) = \sqrt{N} \frac{\partial \ve{C}(\ve{\theta}) }{\partial [\ve{\theta}]_i} \ve{\tilde{g}} .
\end{equation}
The filter spectrum points
\begin{align}
\ve{\tilde{h}}_{k} &=\trans{[H_{k-L^-_kN}, H_{k-(L^-_k-1)N}, \dots H_{k+L^+_kN}]},
\end{align}
separated by $\omega_s$, experience aliasing since they are combined by the multiplication of $\ve{\xi}_i(\ve{\theta}) \circ \ve{\tilde{h}}$ with $\ve{A}$, where the corresponding coefficients are
\begin{align}
\ve{\xi}_{k,i}(\ve{\theta}) &= \Big[ [\ve{\xi}_i(\ve{\theta})]_{k-L^-_kN+\frac{K}{2}+1}, \ve{\xi}_i(\ve{\theta})]_{k-(L^-_k-1)N+\frac{K}{2}+1}, \nonumber \\
&\quad \quad \dots, [\ve{\xi}_i(\ve{\theta})]_{k+L^+_kN+\frac{K}{2}+1} \Big]. \label{eq:xi_mutual_aliasing}
\end{align}
Thus, the compact FIM representation
\begin{equation}\label{eq:FIM:compact:receive:spectrum}
[\ve{\bar{J}}_F(\ve{\theta})]_{ij} = 2 \mathrm{Re} \left\{\sum\limits_{k = -\frac{N}{2}}^{\frac{N}{2}-1} \frac{\ve{\tilde{h}}_k^\mathrm{H} \ve{\xi}_{k,i}^*(\ve{\theta}) \trans{\ve{\xi}_{k,j}}(\ve{\theta}) \ve{\tilde{h}}_k}{\ve{\tilde{h}}_k^\mathrm{H} \ve{\tilde{h}}_k}\right\}
\end{equation}
is obtained. Note that in \eqref{eq:xi_mutual_aliasing} the summand $\frac{K}{2}+1$ assures a correct indexing of the vector $\ve{\xi}_i(\ve{\theta})$. In contrast to \eqref{eq:objective:function:quadratic:transmitter}, eq. \eqref{eq:FIM:compact:receive:spectrum} allows us to characterize the objective function by
\begin{equation}\label{eq:objective:function:quadratic:receiver}
\mathrm{tr}(\ve{M}'\ve{\bar{J}}_D) = \sum\limits_{k \in \mathcal{N}} \frac{\ve{\tilde{h}}_k^\mathrm{H} \ve{\Delta}_k \ve{\tilde{h}}_k}{\ve{\tilde{h}}_k^\mathrm{H} \ve{\tilde{h}}_k}, 
\end{equation}
with
\begin{equation}
\ve{\Delta}_k = \sum_{i=1}^{2} \sum_{j=1}^{2} [\ve{M}']_{ji}\E_{\ve{\theta}}\left[ \ve{\xi}_{k,i}^* \trans{\ve{\xi}_{k,j}} + \ve{\xi}_{k,j}^* \trans{\ve{\xi}_{k,i}} \right].
\end{equation}
Note that \eqref{eq:objective:function:quadratic:transmitter} emphasizes the quadratic dependence of the cost function $\mathrm{tr}(\ve{M}'\ve{\bar{J}}_D) $ on the transmit spectrum $\ve{\tilde{g}}$, while \eqref{eq:objective:function:quadratic:receiver} is written as a function of the receive filter $\ve{\tilde{h}}$.
\section{Symmetry of the EFIM}
\label{sec:FIM_Symmetry_Properties}
The EFIM exhibits symmetry concerning time and frequency reversal of the two filters $\ve{\tilde{g}}$ and $\ve{\tilde{h}}$. We will see later that this observation can be used to accelerate the optimization algorithm solving the transceiver design problem \eqref{eq:optimization:problem:efim}.
\subsection{EFIM with Mirrored Transceive Filter Spectra}
Define the mirroring matrix
\begin{equation}
\ve{\Pi} = \left[ 
\begin{array}{cccc}
0 & \dots & 0 & 1 \\
\vdots & \iddots & 1  & 0 \\
0 & \iddots & \iddots & \vdots  \\
1 & 0 & \dots & 0 
\end{array}
\right] \in \fieldR^{K \times K},
\end{equation}
which obeys
\begin{align}
\ve{\Pi}^{\mathrm{H}} =\ve{\Pi}^{-1}= \ve{\Pi},
\end{align}
such that the frequency mirroring operations $X(-\omega)$ and $H(-\omega)$ can be denoted as $\ve{\Pi} \ve{\tilde{g}}$ and $\ve{\Pi}\ve{H}\ve{\Pi}$, respectively. Writing $\ve{\bar{J}}_F(\ve{\theta})\big|_{\ve{\tilde{g}},\ve{\tilde{h}}}$ to denote the FIM with transmit $\ve{\tilde{g}}$ and receive filter spectrum $\ve{\tilde{h}}$, we find the FIM entries after mirroring the transceive filters along the frequency axis
\begin{align}\label{eq:FIM:entries:mirrored}
&\Big[\ve{\bar{J}}_F(\ve{\theta})\big|_{\ve{\Pi\tilde{g}},\ve{\Pi\tilde{h}}} \Big]_{ij} =2N\mathrm{Re} \Bigg\{ \herm{\ve{\tilde{g}}} \ve{\Pi} \frac{\partial \herm{\ve{C}}(\ve{\theta}) }{\partial [\ve{\theta}]_i} \ve{\Pi} \herm{\ve{H}} \ve{\Pi} \herm{\ve{A}} \ve{\Pi} \ve{\Omega}_{\ve{\eta}}^{-1} \cdot \nonumber \\ & \quad\quad\quad\quad\quad\quad\quad\quad\quad\quad\quad\quad \ve{\Pi A}  \ve{\Pi H\Pi} \frac{\partial  \ve{C}(\ve{\theta}) }{\partial [\ve{\theta}]_j} \ve{\Pi} \ve{\tilde{g}}  \Bigg\} \nonumber \\
&= 2N\mathrm{Re} \Bigg\{ \herm{\ve{\tilde{g}}} \ve{\Pi}\frac{\partial \herm{\ve{C}}(\ve{\theta}) }{\partial [\ve{\theta}]_i} \ve{\Pi} \herm{\ve{H}} \herm{\ve{A}} \ve{\Omega}_{\ve{\eta}}^{-1}  \ve{A}  \ve{H\Pi} \frac{\partial  \ve{C}(\ve{\theta}) }{\partial [\ve{\theta}]_j} \ve{\Pi} \ve{\tilde{g}} \Bigg\}.
\end{align}
It can be seen that \eqref{eq:FIM:entries:mirrored} contains the mirrored matrix
\begin{align}
\ve{\Pi}\frac{\partial  \ve{C}(\ve{\theta}) }{\partial [\ve{\theta}]_i}\ve{\Pi} &= \frac{\partial}{\partial [\ve{\theta}]_i} \left(\gamma \ve{\Pi}\ve{D}(\nu)\ve{T}(\tau)\ve{\Pi}\right)\notag\\
&= \frac{\partial}{\partial [\ve{\theta}]_i}\left(\gamma\ve{\Pi}\ve{D}(\nu)\ve{\Pi}\ve{\Pi}\ve{T}(\tau)\ve{\Pi}\right).
\end{align}

With the structure of the time-delay matrix \eqref{eq:tau:matrix:entries}, we have
\begin{equation}
\ve{\Pi} \ve{T}(\tau) \ve{\Pi} \approx \ve{T}(-\tau).
\end{equation}
Note that by definition \eqref{eq:tau:matrix:entries} the matrix $\ve{T}(\tau)$ is not perfectly symmetric as the frequency interval ranges from $-\frac{K}{2} \omega_0$ to $(\frac{K}{2}-1)\omega_0$. However, for large $K$ the impact of the single asymmetric frequency point $-\frac{K}{2} \omega_0$ is negligible. Using  the symmetry of the Doppler coefficients
\begin{equation}
d_{-z}(\nu) = \frac{\sin(\pi T_0 \nu + \pi z)}{\pi T_0 \nu + \pi z} =  d_z(-\nu),
\end{equation}
we obtain
\begin{equation}
\ve{\Pi} \ve{D}(\nu) \ve{\Pi} = \ve{D}(-\nu).
\end{equation}
Thus,
\begin{equation}
\ve{\Pi}\frac{\partial  \ve{C}(\ve{\theta}) }{\partial [\ve{\theta}]_i}\ve{\Pi} \approx \frac{\partial  }{\partial [\ve{\theta}]_i} \ve{C}(-\boldsymbol{\theta}) = -\frac{\partial  \ve{C}(\ve{\theta}') }{\partial [\ve{\theta}']_i}\bigg|_{\ve{\theta}' = -\ve{\theta}},
\end{equation}
showing that applying the mirroring operation to both filters corresponds to a reversal of the unknown parameters, i.e.,
\begin{equation} \label{eq:Mirror_FIM}
\Big[\ve{\bar{J}}_F(\ve{\theta})\big|_{\ve{\Pi\tilde{g}},\ve{\Pi\tilde{h}}} \Big]_{ij} \approx \Big[\ve{\bar{J}}_F(-\ve{\theta})\big|_{\ve{\tilde{g}},\ve{\tilde{h}}} \Big]_{ij}.
\end{equation}
The time-delay and Doppler shift are assumed to be normally distributed and thus have symmetric distributions, i.e.,
\begin{equation}
p(\ve{\theta}) = p(-\ve{\theta}).
\end{equation}
In general, mirroring a function when averaging with respect to a symmetric distribution does not influence the expected value and thus we obtain the approximate identity 
\begin{equation}\label{eq:EFIM:mirrored}
\ve{\bar{J}}_D|_{\ve{\tilde{g}},\ve{\tilde{h}}} \approx \ve{\bar{J}}_D|_{\ve{\Pi\tilde{g}},\ve{\Pi\tilde{h}}}.
\end{equation}
Note, that using the properties of the Fourier transform, the same result is obtained when mirroring the transmit signal and received signal in the time domain.
\subsection{Off-Diagonal Elements of the EFIM with Symmetric Filters}
\label{ss:Non_Diagonal_Symmetric}
We will show in the following that for symmetric transceiver filters the non-diagonal elements of the FIM vanish. Due to the symmetry and by the properties of the Fourier transform, $\ve{\tilde{g}}, \ve{\tilde{h}}$ become real-valued. Therefore, we have
\begin{align}
\left[\ve{\bar{J}}_F(\ve{\theta})\right]_{ij} &= 2\mathrm{Re} \left\{ \herm{\ve{\tilde{g}}} \ve{\Phi}_{ij}(\ve{\theta}) \ve{\tilde{g}} \right\} \nonumber \\ &=  \herm{\ve{\tilde{g}}} \ve{\Phi}_{ij}(\ve{\theta}) \ve{\tilde{g}} + \trans{\ve{\tilde{g}}} \ve{\Phi}^*_{ij}(\ve{\theta}) \ve{\tilde{g}}^*  \nonumber \\
&=  \herm{\ve{\tilde{g}}} \left(\ve{\Phi}_{ij}(\ve{\theta}) + \ve{\Phi}^*_{ij}(\ve{\theta}) \right) \ve{\tilde{g}}.
\end{align}
It is straightforward to show that the following properties hold for the complex conjugates
\begin{align}
\ve{T}^*(\tau) &= \ve{T}(-\tau), \\
\ve{D}^*(\nu) &= \ve{D}(\nu).
\end{align}
For the off-diagonal elements of the FIM, the matrices $\ve{\Phi}_{ij}(\ve{\theta})$ contain exactly one partial derivative with respect to $\tau$ and one with respect to $\nu$. Hence, for $i\neq j$ 
\begin{equation} \label{eq:FIM_Non_Diagonal}
\left[\ve{\bar{J}}_F(\ve{\theta})\right]_{ij} = \herm{\ve{\tilde{g}}} \big(\ve{\Phi}_{ij}(\tau,\nu) - \ve{\Phi}_{ij}(-\tau,\nu) \big) \ve{\tilde{g}}. 
\end{equation}
Due to the symmetry of the time-delay distribution $p(\tau)=p(-\tau)$, taking the expected value of \eqref{eq:FIM_Non_Diagonal}, the off-diagonal elements of the EFIM $\ve{\bar{J}}_D$ become
\begin{equation}
[\ve{\bar{J}}_D]_{12} = [\ve{\bar{J}}_D]_{21} = 0.
\end{equation}
\subsection{Compact Cost Functions}
\label{ss:Symmetric_Optimization}
With \eqref{eq:EFIM:mirrored}, the reversal of the transmit and the receive filter in the frequency domain does not significantly change the elements of the EFIM if the parameters are distributed symmetrically. Therefore, we will simplify the optimization problem \eqref{eq:optimization:problem:efim} by exclusively considering transceive filters $\ve{\tilde{g}}$ and $\ve{\tilde{h}}$ which have symmetric impulse responses and spectra. Note, that due to the properties of the Fourier transform and filter symmetry, both filters are real-valued and optimization is performed by adapting half of the spectrum.

We split the transmit signal
\begin{equation}\label{transmit:filter:spectrum:splitted}
\ve{\tilde{g}} = \big[ 0 \quad \ve{\tilde{g}}_\mathrm{r}^{\mathrm{T}} \quad \tilde{g}_0 \quad \ve{\tilde{g}}_{\mathrm{f},\mathrm{r}}^{\mathrm{T}}  \big]^{\mathrm{T}},
\end{equation}
with the one-sided transmit spectrum $\ve{\tilde{g}}_\mathrm{r} \in \fieldR^{\frac{K}{2}-1}$ and  $\ve{\tilde{g}}_{\mathrm{f},\mathrm{r}} = \ve{\Pi} \ve{\tilde{g}}_\mathrm{r}$. Note that the first element in \eqref{transmit:filter:spectrum:splitted} has been set to zero, i.e., $G_{-\frac{K}{2}}=0$, to obtain a fully symmetric description. This allows us to rewrite the cost function \eqref{eq:objective:function:quadratic:transmitter} as
\begin{align}
\ve{\tilde{g}}^\mathrm{H} \ve{\Phi} \ve{\tilde{g}} &= \left[ 0 \quad \ve{\tilde{g}}_{\mathrm{r}}^{\mathrm{T}} \quad \tilde{g}_0 \quad \ve{\tilde{g}}_{\mathrm{f},\mathrm{r}}^{\mathrm{T}}  \right] \ve{\phi} \left[ 
\begin{array}{c}
0 \\
\ve{\tilde{g}}_{\mathrm{r}} \\
\tilde{g}_0  \\
\ve{\tilde{g}}_{\mathrm{f},\mathrm{r}}
\end{array}
\right] \nonumber \\
&=\big[\ve{\tilde{g}}_{\mathrm{r}}^{\mathrm{T}} \quad \tilde{g}_0 \quad \ve{\tilde{g}}_{\mathrm{f},\mathrm{r}}^{\mathrm{T}}  \big] \left[ 
\begin{array}{ccc}
\ve{\phi}_{11} & \ve{\phi}_2  & \ve{\phi}_{12} \\
\ve{\phi}_1^\mathrm{T} & \phi_0 & \ve{\phi}_3^\mathrm{T}  \\
\ve{\phi}_{21} & \ve{\phi}_4 & \ve{\phi}_{22} 
\end{array}
\right] \left[ 
\begin{array}{c}
\ve{\tilde{g}}_{\mathrm{r}} \\
\tilde{g}_0  \\
\ve{\tilde{g}}_{\mathrm{f},\mathrm{r}}
\end{array}
\right] \nonumber \\
&=\big[\ve{\tilde{g}}_{\mathrm{r}}^{\mathrm{T}} \quad \tilde{g}_0 \big] \left[ \begin{array}{cc}
\ve{\phi}_{r} & \ve{\phi}_2  + \ve{\Pi}\ve{\phi}_4 \\
(\ve{\phi}_1 + \ve{\Pi}\ve{\phi}_3)^\mathrm{T} & \phi_0 \end{array}
\right] \left[ 
\begin{array}{c}
\ve{\tilde{g}}_{\mathrm{r}} \\
\tilde{g}_0  \end{array}
\right] \label{reduced:quadratic:form:receive:filter},
\end{align}
with the matrices and vectors
\begin{align}
\ve{\phi}_{11}, \ve{\phi}_{12}, \ve{\phi}_{21}, \ve{\phi}_{22} &\in \fieldC^{\left(\frac{K}{2}-1\right) \times \left(\frac{K}{2}-1\right)}, \\
\ve{\phi}_1, \ve{\phi}_2, \ve{\phi}_3, \ve{\phi}_4 &\in \fieldC^{\frac{K}{2}-1}
\end{align}
and a reduced filter matrix
\begin{equation}
\ve{\phi}_{\mathrm{r}} = \ve{\phi}_{11} + \ve{\phi}_{12} \ve{\Pi} + \ve{\Pi} \ve{\phi}_{21} + \ve{\Pi} \ve{\phi}_{22} \ve{\Pi}.
\end{equation}
It can be seen that with symmetric filters the cost function \eqref{eq:objective:function:quadratic:transmitter} can be written in a compact quadratic form \eqref{reduced:quadratic:form:receive:filter}, requiring only one side of the transmit filter spectrum $\ve{\tilde{g}}$. A compact version of \eqref{eq:objective:function:quadratic:receiver} can be found in a similar way by using
\begin{equation}
\ve{\tilde{h}}_k = \ve{\Pi}\ve{\tilde{h}}_{-k}.
\end{equation}
This allows to rewrite the cost function \eqref{eq:objective:function:quadratic:receiver} as
\begin{align}\label{eq:objective:function:quadratic:receiver:symmetric}
\sum\limits_{k = -\frac{N}{2}}^{\frac{N}{2}-1} \frac{\ve{\tilde{h}}_k^\mathrm{H} \ve{\Delta}_k \ve{\tilde{h}}_k}{\ve{\tilde{h}}_k^\mathrm{H} \ve{\tilde{h}}_k} &= \sum\limits_{k =-\frac{N}{2}+1}^{-1} \frac{\ve{\tilde{h}}_k^\mathrm{H} (\ve{\Delta}_k + \ve{\Pi} \ve{\Delta}_{-k} \ve{\Pi}) \ve{\tilde{h}}_k}{\ve{\tilde{h}}_k^\mathrm{H} \ve{\tilde{h}}_k}  \nonumber \\ 
&\quad+\sum_{k \in \left\{-\frac{N}{2}, 0\right\}} \frac{\ve{\tilde{h}}_k^\mathrm{H} \ve{\Delta}_k \ve{\tilde{h}}_k}{\ve{\tilde{h}}_k^\mathrm{H} \ve{\tilde{h}}_k}.
\end{align}
\section{Transceiver Optimization Algorithm}
\label{sec:optimization:algorithm}
In the following we focus on solving the transceiver design problem \eqref{eq:optimization:problem:efim} with the approximate EFIM derived in the previous section. Further, we will present an algorithm which solves the maximization task \eqref{eq:Opt_Approx:frequency:representation}. With the frequency domain representation \eqref{frequency:representation:transmit:filter} and \eqref{frequency:representation:receive:filter:simplified}, the optimization problem \eqref{eq:optimization:problem:efim} can be stated as
\begin{equation}\label{eq:Opt_Approx:frequency:representation}
\underset{\ve{\tilde{g}}, \ve{\tilde{h}}}{\arg\max}\, \mathrm{tr} \left(\boldsymbol{M}' \ve{\bar{J}}_D|_{\ve{\tilde{g}},\ve{\tilde{h}}}\right) \quad \mathrm{s.t.}\quad\herm{\ve{\tilde{g}}} \ve{\tilde{g}} \leq P_T.
\end{equation}
Note that in contrast to \eqref{eq:optimization:problem:efim}, where the exact EFIM is used, the problem is now formulated based on the EFIM approximation $\ve{\bar{J}}_D$ that has been introduced in Section \ref{sec:approximation:fim}.

We first focus on the problem of optimizing the transmit signal $\boldsymbol{\tilde{g}}$ for a given receive filter $\boldsymbol{\tilde{h}}$
\begin{align}\label{optimization:alternating:transmitter}
\underset{\boldsymbol{\tilde{g}}}{\arg\max} \, \mathrm{tr} \left(\boldsymbol{M}' \ve{\bar{J}}_D|_{\ve{\tilde{g}},\ve{\tilde{h}}}\right)
\quad\mathrm{s.t.}\quad
\herm{\boldsymbol{\tilde{g}}} \boldsymbol{\tilde{g}} \leq P_T,
\end{align}
where the cost function is expressed by \eqref{eq:objective:function:quadratic:transmitter}. It is well known that the solution to such a problem is the Eigenvector $\boldsymbol{\gamma}_1$ of the matrix $\boldsymbol{\Phi}$ corresponding to the largest Eigenvalue \cite{Golub73}.

On the other hand, for a constant transmit signal $\boldsymbol{\tilde{g}}$, the optimization problem reads as
\begin{align}\label{optimization:alternating:receiver}
\underset{\boldsymbol{\tilde{h}}}{\max}\quad \mathrm{tr} \left(\boldsymbol{M}' \ve{\bar{J}}_D|_{\ve{\tilde{g}},\ve{\tilde{h}}}\right).
\end{align}
Having a closer look at the filter representation of the objective function \eqref{eq:objective:function:quadratic:receiver}, it is straightforward to prove that two different vectors $\boldsymbol{\tilde{h}}_{k_1}$ and $\boldsymbol{\tilde{h}}_{k_2}$, $k_1 \neq k_2$ consist of two disjoint sets of filter coefficients $H_k$. Therefore problem \eqref{optimization:alternating:receiver} is solved by independently maximizing each term of the sum \eqref{eq:objective:function:quadratic:receiver:symmetric}, which results in choosing $\boldsymbol{\tilde{h}}_k = \boldsymbol{\delta}_{k,1}$ where $\boldsymbol{\delta}_{k,1}$ is the Eigenvector of $\boldsymbol{\Delta}_k$ corresponding to the largest Eigenvalue.

With the two separate optimization tasks \eqref{optimization:alternating:transmitter} and \eqref{optimization:alternating:receiver} for the transmit and the receive filter, we suggest the following iterative algorithm to find the solution to the design problem in \eqref{eq:Opt_Approx:frequency:representation}. The algorithm is initialized with filters $\ve{\tilde{g}}_{\mathrm{init}}$ and $\ve{\tilde{h}}_{\mathrm{init}}$ that can either be chosen randomly or set to initial prototype versions. In an alternating way, in the $i$-th iteration the analog transmit filter $\boldsymbol{\tilde{g}}$ is optimized by solving \eqref{optimization:alternating:transmitter} for a fixed receive filter $\boldsymbol{\tilde{h}}^{i-1}$. The solution $\boldsymbol{\tilde{g}}^i$ is then used to improve the solution for $\boldsymbol{\tilde{h}}$ with \eqref{optimization:alternating:receiver}. This approach ensures that the value of the objective function increases with each iteration. If the relative performance gain is less than a threshold value $\epsilon$, the algorithm terminates and outputs the solutions $\boldsymbol{\tilde{g}}^\star$ and $\boldsymbol{\tilde{h}}^\star$. Note that such an alternating approach does not necessarily find the globally optimum solution. However, simulations show (see Sections \ref{sec:optimization:results} and \ref{sec:simulation:results}) that the procedure outputs a transceiver design which provides significant performance gains. For all tested scenarios, the suggested algorithm converges quickly and returns a filter pair with favorable design after a few steps.
\section{Reference Systems}
\label{sec:reference:systems}
To provide a comparison to commonly used transceive filters, we introduce two reference designs which will serve as a benchmark for the optimized solutions. We use rectangular pulses as employed in Global Navigation Satellite Systems (GNSS) \cite{Misra11} such as the American Global Positioning System (GPS), and linear frequency modulation (LFM) pulses that find application in long-range, high-resolution radar systems \cite[p. 88]{cook1993}. For further details about LFM pulses, the reader is referred to \cite{Klauder60}. 

Both reference signals $\breve{x}(t)$ can be described by the periodic repetition of a transmit pulse $g(t)$ in intervals of $T_0$. Having available the spectrum $G(\omega)$ of the reference pulse, the transmit filter Fourier coefficients are simply obtained by the evaluation of the discrete spectrum points $G(k\omega_0)$ and normalizing to $\herm{\boldsymbol{\tilde{g}}}_{\mathrm{ref}}\boldsymbol{\tilde{g}}_{\mathrm{ref}} = P_T$. The analog reference receive filter $\ve{\tilde{h}}_{\mathrm{ref}}$ is an ideal low-pass filter with bandwidth $B_{\mathrm{ref}}=f_s$ which complies with the sampling theorem. 
To visualize the possible performance gains, in the following, we consider the relative performance measures
\begin{equation}\label{def:relative:performance}
\chi_{\tau/\nu} = 10 \log \left( \frac{\big[\boldsymbol{J}_D^{-1}|_{\ve{\tilde{g}}_\mathrm{ref},\ve{\tilde{h}}_\mathrm{ref}}\big]_{11/22}}{\big[\boldsymbol{J}_D^{-1}|_{\ve{\tilde{g}},\ve{\tilde{h}}}\big]_{11/22}} \right),
\end{equation}
which characterize the information gain in the data influenced part of the BCRLB \eqref{eq:BCRLB} and therefore characterize the factor by which the transmit power of the optimized system could be reduced while maintaining the performance level of the reference system. Note that \eqref{def:relative:performance} is computed using the exact EFIM \eqref{eq:EFIM} with entries \eqref{eq:FIM_11} -- \eqref{eq:FIM_22}. The reference systems are configured such that they use the same transmit power $P_T$ as the optimized system.
\subsection{Rectangular Phase Code Modulation}
The rectangular prototype pulse has duration $T= 2T_s$ and its Fourier transform can be denoted
\begin{align}
g^{(\infty)}_{\mathrm{Rect}}(t) &=  \mathrm{rect}\left( \frac{t}{T} \right), \\
G^{(\infty)}_{\mathrm{Rect}}(\omega) &= \frac{2\sin\left(\omega \frac{T}{2}\right)}{\omega}.
\end{align}
Under the assumption that the pulse is ideally band-limited to a two-sided bandwidth $B= f_s$ at the transmitter, we obtain
\begin{align}
G_{\mathrm{Rect}}(\omega) &= \left\{ \begin{array}{lll}
\frac{2\sin\left(\omega \frac{T}{2}\right)}{\omega}, & -\pi B \leq \omega < \pi B \\
0, & \mathrm{otherwise}
\end{array} \right. , \\
g_\mathrm{Rect}(t) &= \int\limits g^{(\infty)}_\mathrm{Rect}(\psi) \frac{\sin\big( \pi B (t-\psi)\big)}{\pi (t-\psi)} \mathrm{d} \psi.
\end{align}
The rectangular phase code (RPC) is then generated by convolving the prototype rectangular pulse with a code sequence
\begin{equation}\label{def:reference:rpc}
g_\mathrm{RPC}(t) = g_\mathrm{Rect}(t) * \beta(t),
\end{equation}
where
\begin{equation}\beta(t) = \sum_{m=0}^{M-1} \beta_m \delta(t-mT)
\end{equation}
for some known binary sequence $\beta_m \in \{-1,1\}$.
\subsection{Linear Frequency Modulation (LFM)}
The LFM reference signal consists of an envelope $a_{\mathrm{LFM}}(t)$ and a sinusoid with linearly increasing frequency $w_{\mathrm{LFM}}(t)$. For the case of a rectangular envelope, we have
\begin{align}
a_{\mathrm{LFM}}(t) &=  \mathrm{rect}\left( \frac{t}{T} \right), \\
w_{\mathrm{LFM}}(t) &= \e^{\mathrm{j} \frac{\mu t^2}{2}},\label{LFM:chirp}
\end{align}
with pulse duration $T$ and frequency slope $\mu$. In the literature, \eqref{LFM:chirp} is often referred to as chirp pulse. The frequency of the sinusoid is increasing linearly in time
\begin{equation}
\omega = \frac{\mathrm{d} \, \angle(w_{\mathrm{LFM}}(t))}{\mathrm{d} t} = \mu t.
\end{equation}
The LFM transmit pulse and its spectrum are \cite[p. 138]{cook1993}
\begin{align}
g_{\mathrm{LFM}}(t) &= a_{\mathrm{LFM}}(t) w_{\mathrm{LFM}}(t) \label{def:LMF:pulse:time:domain}\\
G_{\mathrm{LFM}}(\omega) &= \frac12 \sqrt{\frac{\pi}{\mu}} \e^{-\mathrm{j} \frac{\omega^2}{2\mu}}\Bigg[Z\left(\sqrt{\frac{1}{\pi\mu}}\omega + \sqrt{\frac{\mu T^2}{4 \pi}}\right) + \nonumber \\ & \quad\quad\quad \quad\quad\quad\quad Z\left(-\sqrt{\frac{1}{\pi\mu}}\omega + \sqrt{\frac{\mu T^2}{4 \pi}}\right)\Bigg],
\end{align}
with the complex Fresnel integral $Z(u) = \int_{0}^{u} \e^{\mathrm{j}\pi \frac{a^2}{2}}\mathrm{d}a$.
The two-sided bandwidth of this pulse is approximately $B_\mathrm{LFM}=\frac{\mu T}{2\pi}$. For the reference system, we use a pulse length of $T=T_0$. The bandwidth $B_\mathrm{LFM} = \frac{2f_s}{3}$ uses the full frequency band  without violating the sampling theorem and yields $\mu = \frac{4\pi N}{3T_0^2}$.
\section{Optimization Results}
\label{sec:optimization:results}
\subsection{Performance - Pareto-optimal Transceiver Region}
In general there exists a trade-off between the estimation accuracy of the time-delay and the Doppler shift.  As such, for the presented transceiver optimization framework we examine the Pareto-optimal set $\mathcal{P} = \{ (g(t), h(t)) \}$ of transmit and receive filters, for which the estimation of one parameter cannot be improved through the filter design without reducing the accuracy of the other parameter. We define the set $\mathcal{G}$ of relative performance gains \eqref{def:relative:performance} which are obtained by filters on the boundary of the Pareto-optimal region $\mathcal{P}$
\begin{equation}
\mathcal{G} = \Big\{ (\chi_\tau, \chi_\nu )\big|_{\mathcal{P}} \Big\}.
\end{equation}
The set $\mathcal{G}$ can be approximated by solving the transceiver optimization problem \eqref{eq:Opt_Approx:frequency:representation} over all positive definite weighting matrices $\boldsymbol{M}'$ and computing the corresponding EFIM values \eqref{eq:EFIM}. In Section \ref{ss:Non_Diagonal_Symmetric} it has been shown that for the examined system the non-diagonal elements of the EFIM $\boldsymbol{\bar{J}}_D$ vanish such that it is sufficient to consider the diagonal weighting matrices
\begin{equation}
\boldsymbol{M}' = \mathrm{diag} (\alpha, 1-\alpha),
\end{equation}
with $\alpha\in [0,1]$.
\begin{figure}[!htbp]
	\centering
	\includegraphics{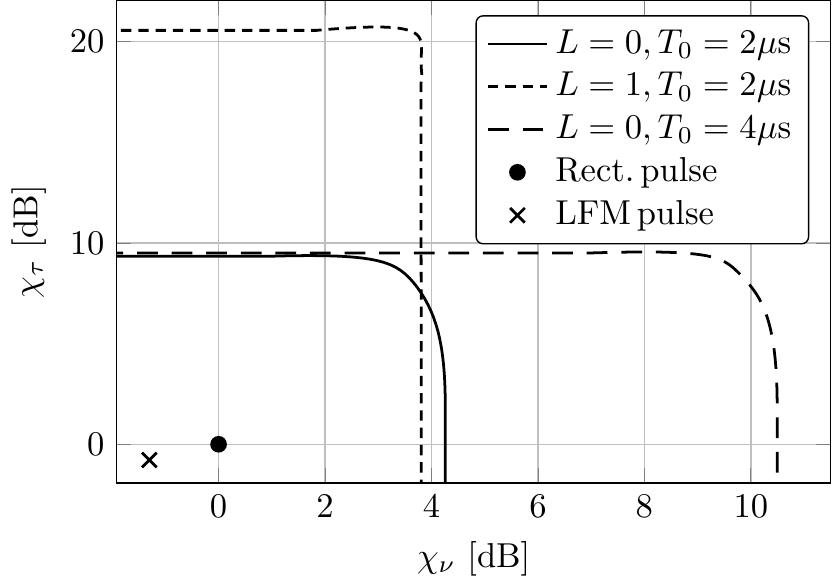}
	\caption{Pareto regions for bandwidths $B=(2L+1)f_s$ and signal periods $T_0$ with $f_s = 25 \mathrm{MHz}$}
	\label{fig:Pareto_Different_Bandwidths}
\end{figure}
\subsubsection{Variable Bandwidth with Fixed Sampling Rate}
Fig. \ref{fig:Pareto_Different_Bandwidths} shows the Pareto boundary for different bandwidths $B = (2L+1) f_s$ and signal periods $T_0$, where $f_s = 25 \mathrm{MHz}$ has been used. Here, the rectangular transmit waveform \eqref{def:reference:rpc} with $T_0 = 2 \mu \mathrm{s}$ is used as the reference in \eqref{def:relative:performance} and the variances of the random unknown parameters \eqref{covariance:matrix:prior} are set to $\sigma_\tau = 1 \mathrm{ns}, \sigma_\nu = 5 \mathrm{kHz}$. Fig. \ref{fig:Pareto_Different_Bandwidths} has been created by solving \eqref{eq:optimization:problem:efim} using the approximated FIM \eqref{eq:Doppler_Approximation_Fim_Entries} for symmetric signals under all weightings $\alpha$ and computing the corresponding exact values $\chi_{\tau/\nu}$ \eqref{def:relative:performance}. Note that the system $L=0, T_0 = 4 \mu \mathrm{s}$ operates at half the transmit power $\frac{P_T}{2}$ such that the total transmitted energy $P_T T_0$ is the same as in the reference system. It can be observed that for the estimation of the time-delay $\tau$, increasing the bandwidth $B$ leads to a substantial performance gain, while for the estimation of the Doppler, an extension of the time-spread $T_0$ improves the achievable accuracy. The fact that the Pareto region appears to be larger for the Doppler parameter when the bandwidth is lower can be explained by the inaccuracies that result from the approximation of the EFIM used in \eqref{eq:Opt_Approx:frequency:representation}. We see that for essentially any choice of the weighting matrix, the optimized design provides significant gains in estimation performance compared with the prototypes. We also note that the nearly rectangular Pareto regions indicate that, with optimized filter designs, the trade-off between delay and Doppler estimation performance is relatively small.
\subsubsection{Fixed Bandwidth with Variable Sampling Rate}
We have seen that optimized filters have the potential to increase the estimation performance of delay-Doppler estimation methods. We now investigate the estimation performance for a fixed transmit bandwidth $B = 25 \mathrm{MHz}$, a signal period $T_0 = 2.4 \mu \mathrm{s}$ and different sampling frequencies $f_s = \frac{B}{2L+1}$.
\begin{figure}[!htbp]
	\centering
	\includegraphics{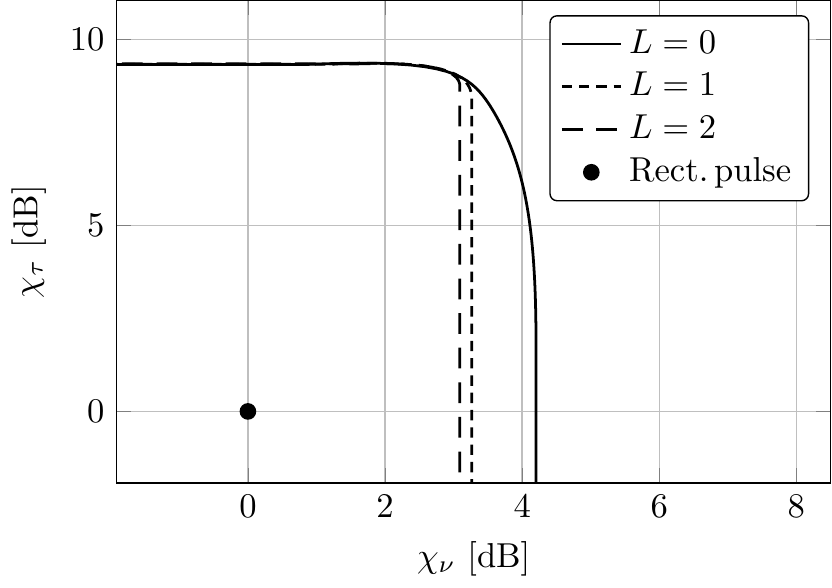}
	\caption{Pareto regions for rates $f_s = \frac{B}{2L+1}$ with $B = 25 \mathrm{MHz}$}
	\label{fig:Pareto_Different_Rates}
\end{figure}
Fig. \ref{fig:Pareto_Different_Rates} shows the Pareto regions of the optimized waveforms with respect to a rectangular signal, all with the same bandwidth of $B=25\mathrm{MHz}$. Hence, the number of Fourier coefficients $K=BT_0$ is the same for all systems. Note that the sampling rate for the reference system is held constant, while the sampling rate and the number of samples $N$ of the optimized system decrease with increasing $L$. This indicates that although lower sampling rates are used, the optimized waveform design can still provide high estimation accuracy. In particular, this means that, if the bandwidth $B$ is fixed and the transceive filters are chosen according to the proposed design rule, reducing the number of samples inside the constant time interval $T_0$ does not result in an accuracy loss for the estimation of $\tau$. There are two main reasons that explain this phenomenon. First, the time delay estimation accuracy benefits from transmit spectra whose power is concentrated at high frequencies (see, e.g., Fig. \ref{fig:Sig_Opt_Spec_L0_Del}). By violating the design rule $B \leq f_s$, the transmit spectrum can be chosen to feature bandwidths with high frequencies, although the sampling bandwidth is decreased. Second, the power loss induced by the reduced number of samples is compensated by a bandpass structure of the receive filter (see Fig. \ref{fig:Filt_Opt_Spec_L1_TOff}), which decreases the power of the noise accordingly. However, the Doppler estimation suffers slightly from a reduced number of sampling points.
\subsubsection{Comparison of Approximate and Exact EFIM}
In the previous paragraphs, the transmit and receive filter optimization was based on the approximated EFIM. In this section, we investigate the accuracy of the noise covariance and Doppler approximation. We show that evaluating the approximated and exact EFIM for the optimized systems leads to similar Pareto regions.
\begin{figure}[!htbp]
	\centering
	\includegraphics{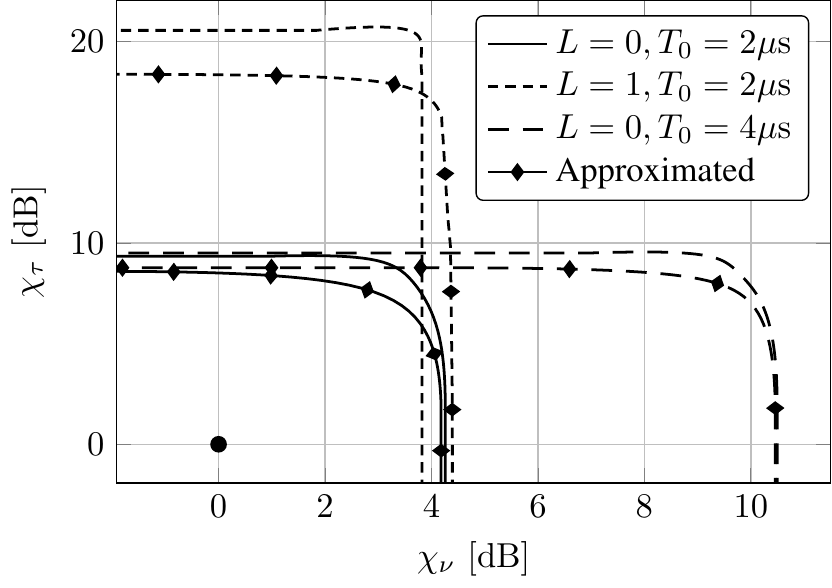}
	\caption{Exact and Approximated Pareto regions for bandwidths $B=(2L+1)f_s$ and signal periods $T_0$ with $f_s = 25 \mathrm{MHz}$}
	\label{fig:Pareto_Approx_Vs_Real}
\end{figure}
Figure \ref{fig:Pareto_Approx_Vs_Real} shows the Pareto-optimal region of the approximated and exact Fisher information matrix, for the case of a rectangular reference signal \eqref{def:reference:rpc}. Fig. \ref{fig:Pareto_Approx_Vs_Real} was created by solving the transceiver optimization problem \eqref{eq:optimization:problem:efim} using the approximated EFIM \eqref{eq:objective:function:quadratic:transmitter} under all weightings $\alpha$, as illustrated in Section \ref{sec:optimization:algorithm}. The optimized filters were then used to compute the corresponding exact performance gains $\chi_{\tau/\nu}$ \eqref{def:relative:performance} using \eqref{eq:FIM_11}-\eqref{eq:FIM_22}. For comparison, approximated values $\bar{\chi}_{\tau/\nu}$ have been calculated using \eqref{eq:Doppler_Approximation_Fim_Entries}. The system parameters were chosen equal to the earlier simulations, i.e., $f_s = 25 \mathrm{MHz}$ and $B = (2L+1) f_s$. The parameter variances are $\sigma_\tau = 1 \mathrm{ns}$, $\sigma_\nu = 5 \mathrm{kHz}$. It can be observed that the approximated bound lies close to the exact bound, with a gap of maximum $2\mathrm{dB}$ and exhibits a similar shape. Note that the optimization in Section \ref{sec:optimization:algorithm} is performed by using the approximated information measure \eqref{eq:objective:function:quadratic:transmitter}. Therefore higher gains might be achievable when using a more accurate representation of the exact EFIM.
\subsection{Design - Optimized Transmit and Receive Filters}
To illustrate transceiver designs optimized for the estimation of $\tau$ and $\nu$, we visualize time and frequency domain representations of the analog receive and transmit filters obtained with the presented optimization algorithm. For the same configuration as in Fig. \ref{fig:Pareto_Different_Bandwidths} and a signal period of $T_0 = 2\mu \mathrm{s}$ we first investigate the case where the sampling theorem is satisfied, i.e., $B = f_s$. Note that under these conditions the optimum receive filter is an ideal rectangular low-pass filter with two-sided bandwidth $B_\mathrm{LP} = f_s$.

\begin{figure}[!htbp]
	\centering
	\subfloat[Frequency domain ($\alpha=1$) \label{fig:Sig_Opt_Spec_L0_Del}]{\includegraphics{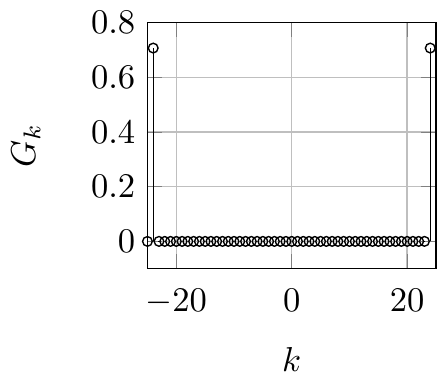}}
	\subfloat[Time domain ($\alpha=0$) \label{fig:Sig_Opt_Time_L0_Dop}]{\includegraphics{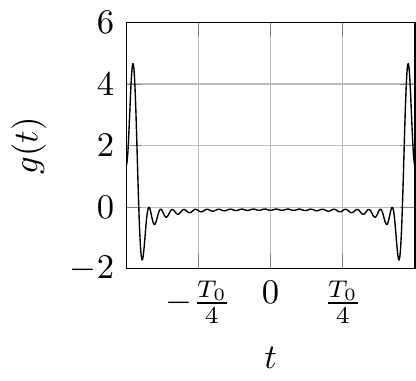}} \\
	\subfloat[Frequency domain ($\alpha=\alpha^\star$) \label{fig:Sig_Opt_Spec_L0_TOff}]{\includegraphics{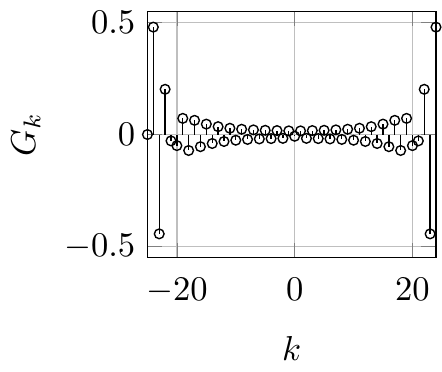}}
	\subfloat[Time domain ($\alpha=\alpha^\star$) \label{fig:Sig_Opt_Time_L0_TOff}]{\includegraphics{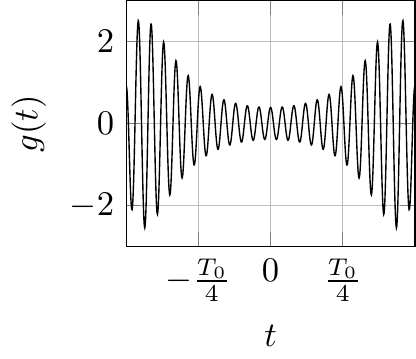}}
	\caption{Optimized transmit filters ($L=0$)}
	\label{fig:Sig_Opt_L0}			
\end{figure}

Fig. \ref{fig:Sig_Opt_Spec_L0_Del} shows the spectrum of the optimized transmit signal in the frequency-domain when full weight is given to the optimization of the delay estimation ($\alpha = 1$).
In contrast Fig. \ref{fig:Sig_Opt_Time_L0_Dop} depicts the result of the transmit filter optimization in the time-domain when the inference of the Doppler is emphasized ($\alpha = 0$). It can be observed that when focusing on the estimation of the delay ($\alpha=1$), the transmit signal power is concentrated at high frequencies (see Fig. \ref{fig:Sig_Opt_Spec_L0_Del}). When enforcing estimation of the Doppler, the signal power is mainly located at the boundaries of the observation interval (see Fig. \ref{fig:Sig_Opt_Time_L0_Dop}). This is because for delay estimation, signal components with high frequencies are more beneficial for the EFIM, since the phase shift, which is introduced by the propagation delay, increases linearly in frequency. On the other hand, for Doppler estimation a signal with wide time-spread is beneficial as the phase shift that is introduced by the Doppler effect increases linearly in time. Using a weighting
\begin{equation}\label{def:equal:weighting}
\alpha^\star =  \arg\underset{0<\alpha<1}{\max} \, \chi_\tau + \chi_\nu, \quad 
\end{equation}
which simultaneously emphasizes both the estimation of the delay and Doppler shift, we obtain the transmit filter depicted in Figs. \ref{fig:Sig_Opt_Spec_L0_TOff} and \ref{fig:Sig_Opt_Time_L0_TOff}. Note that $\alpha^*$ leads to the point on $\mathcal{G}$ at which the sum of both accuracy gains \eqref{def:relative:performance} is maximum, i.e., the most upper right point in the Pareto region $\mathcal{G} $. It can be observed that in this case, the signal power is concentrated at the boundaries of the time and frequency domain. Due to the properties of the Fourier transform, the signal cannot be limited in time and frequency simultaneously, which leads to the observed trade-off between delay and Doppler estimation. Furthermore, it is noticeable that the optimized transmit signal exhibits strong oscillations in the time and frequency domain (Figs. \ref{fig:Sig_Opt_Spec_L0_TOff} and \ref{fig:Sig_Opt_Time_L0_TOff}).
\begin{figure}[!htbp]
	\centering
	\subfloat[Transmit filter spectrum \label{fig:Sig_Opt_Spec_L1_TOff}]{\includegraphics{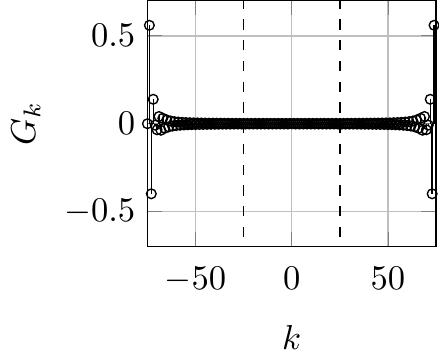}}
	\subfloat[Transmit filter impulse \label{fig:Sig_Opt_Time_L1_TOff}]{\includegraphics{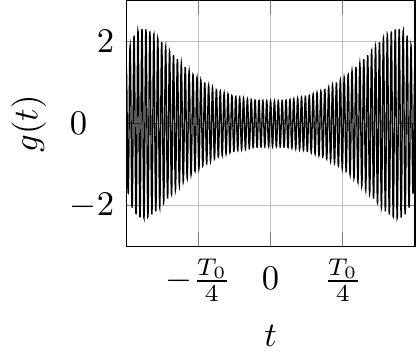}} \\
	\subfloat[Periodic transmit filter spectrum \label{fig:Sig_Opt_SpecRep_L1_TOff}]{\includegraphics{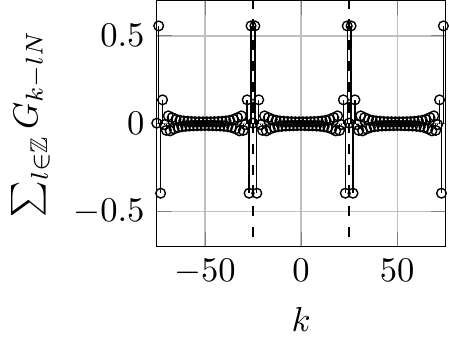}}
	\subfloat[Receive filter spectrum \label{fig:Filt_Opt_Spec_L1_TOff}]{\includegraphics{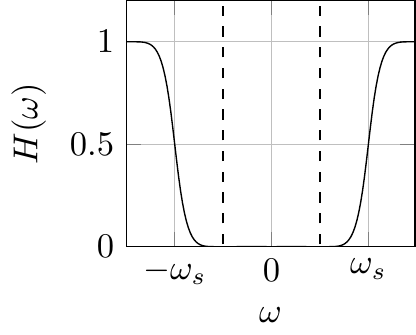}}
	\caption{Optimized transmit and receive filters ($L=1, \alpha=\alpha^\star$)}
	\label{fig:Sig_Opt_L1}			
\end{figure}
Allowing for a higher bandwidth $L=1$, i.e., $B = 3f_s$ and therefore for received signals which violate the sampling theorem, Figs. \ref{fig:Sig_Opt_Spec_L1_TOff} and \ref{fig:Sig_Opt_Time_L1_TOff} show the optimized waveform for the max-sum  weighting $\alpha^{\star}$ \eqref{def:equal:weighting}. The dashed lines visualize the interval $[-\frac{\omega_s}{2}, \frac{\omega_s}{2})$. It can be observed that the optimized transmit filter has a total bandwidth of $B > f_s$ and therefore violates the sampling theorem. However, as this signal shows a bandpass character, the real bandwidth of the signal, i.e., the cumulative portion of the frequencies where the signal exhibits non-zero spectral power density is merely $\omega_s$. Although the total bandwidth of the signal exceeds $\omega_s$, due to its bandpass character the transmit signal does not experience significant aliasing during the sampling process at the receiver. Comparing the sampled signal spectrum in Fig. \ref{fig:Sig_Opt_SpecRep_L1_TOff} with the transmit spectrum in Fig. \ref{fig:Sig_Opt_Spec_L1_TOff}, it can be seen that signal replicas are not distorted. The advantage of the band-pass character of the transmitter is a high sensitivity at the receiver with respect to the propagation delay. Fig. \ref{fig:Filt_Opt_Spec_L1_TOff} displays the optimized receive filter spectrum, which is a bandpass. An interesting observation is that none of the optimized signals uses the full available bandwidth $B$. Note that due to the relatively small prior variances, in general, the full bandwidth and time interval could be used. For higher parameter uncertainties, safety margins at the interval boundaries would occur to prevent a potential power loss from large time or frequency shifts.
\subsection{Complexity - Ambiguity Function}
\label{ss:Ambiguity_Function}
The ambiguity function, defined as the output of a matched filter without Doppler shift compensation
\begin{equation}\label{def:ambiguity:function}
\psi(\boldsymbol{\theta}) = \left| \int_{-\infty}^{\infty} \breve{x}(t) \breve{x}^*(t+\tau) \e^{\mathrm{j}2\pi\nu t} \mathrm{d}t \right|^2,
\end{equation}
is a well-known indicator for analyzing the suitability of transmit waveforms for delay-Doppler estimation\cite[ch. 4]{cook1993}. It can be shown that \eqref{def:ambiguity:function} is directly related to the functional distance of the transmit waveform and its time and frequency-shifted versions. It therefore provides insight about the self-similarity of the signal and its sensitivity with respect to shifts in time and frequency. By discussing the curvature of \eqref{def:ambiguity:function} at $\psi(0, 0) $ and its side-lobe levels, it is possible to establish a measure of the adequateness of a waveform with respect to delay and Doppler estimation \cite{Gu96}.
\subsubsection{MAP Estimation}
In a realistic scenario the attenuation $\gamma$ in the received signal model \eqref{eq:Delay_Doppler_Definition} is deterministic and unknown, so the joint maximum-a-posteriori/maximum-likelihood (JMAP-ML) approach \cite{Yeredor00}
\begin{align}\label{def:jmap:ml}
\begin{bmatrix}
\hat{\gamma}_{\mathrm{ML}}(\boldsymbol{y})\\
\boldsymbol{\hat{\theta}}_{\mathrm{MAP}}(\boldsymbol{y})
\end{bmatrix}
&= \arg\underset{\boldsymbol{\theta},\gamma}{\max} \, \big(\ln p(\boldsymbol{y}|\boldsymbol{\theta},\gamma) + \ln p(\boldsymbol{\theta})\big)
\end{align}
has to be used to estimate the unknown parameters. For the case of Gaussian  noise, the likelihood function becomes
\begin{equation}
p(\boldsymbol{y}|\boldsymbol{\theta},\gamma)  \propto \exp\Big(- \frac{1}{2} \big(\boldsymbol{y} - \boldsymbol{v} (\boldsymbol{\theta},\gamma)\big)^\mathrm{H} \boldsymbol{R}_{\boldsymbol{\eta}}^{-1} \big(\boldsymbol{y} - \boldsymbol{v} (\boldsymbol{\theta},\gamma)\big)\Big).
\end{equation}
Writing
\begin{equation}
\boldsymbol{v}(\boldsymbol{\theta},\gamma) = \gamma \boldsymbol{s}(\boldsymbol{\theta}),
\end{equation}
we find that the solution of \eqref{def:jmap:ml} needs to satisfy
\begin{align}\label{eq:ML_Der}
\frac{\partial }{\partial \gamma^*}\ln p(\boldsymbol{y}|\boldsymbol{\theta},\gamma)&= \herm{\boldsymbol{s}}(\boldsymbol{\theta})\boldsymbol{R}_{\boldsymbol{\eta}}^{-1}\left(\boldsymbol{y} - \gamma \boldsymbol{s}(\boldsymbol{\theta})\right)=0.
\end{align}
Hence, we obtain the estimator for the path gain
\begin{equation}\label{result:ml:gamma}
\hat{\gamma}(\boldsymbol{\theta}) = \frac{\herm{ \boldsymbol{s}}(\boldsymbol{\theta})\boldsymbol{R}_{\boldsymbol{\eta}}^{-1}\boldsymbol{y}}{\herm{ \boldsymbol{s}}(\boldsymbol{\theta})\boldsymbol{R}_{\boldsymbol{\eta}}^{-1}\boldsymbol{s}(\boldsymbol{\theta})},
\end{equation}
as a function of the parameters $\boldsymbol{\theta}$. Substituting \eqref{result:ml:gamma} into \eqref{def:jmap:ml}, we obtain the estimator
\begin{align}\label{def:MAP:Doppler:delay}
&\boldsymbol{\hat{\theta}}_{\mathrm{MAP}}(\boldsymbol{y}) = \nonumber \\
&=\arg \, \underset{\boldsymbol{\theta}}{\max}\, \big(
\ln p(\boldsymbol{y}|\boldsymbol{\theta},\hat{\gamma}(\boldsymbol{\theta})) + \ln p(\boldsymbol{\theta}) \big)\nonumber \\
&= \arg \, \underset{\boldsymbol{\theta}}{\max}\, \bigg(-\big(\boldsymbol{y} - \hat{\gamma}(\boldsymbol{\theta}) \boldsymbol{s}(\boldsymbol{\theta})\big)^\mathrm{H} \boldsymbol{R}_{\boldsymbol{\eta}}^{-1} \big(\boldsymbol{y} - \hat{\gamma}(\boldsymbol{\theta}) \boldsymbol{s}(\boldsymbol{\theta})\big) \nonumber \\ 
& \quad \quad \quad \quad \quad \quad  - \frac{(\tau-\mu_\tau)^2}{\sigma_\tau^2} -\frac{(\nu-\mu_\nu)^2}{\sigma_\nu^2} \bigg)\nonumber \\
&= \arg \, \underset{\boldsymbol{\theta}}{\max}\, f_\mathrm{MAP}(\boldsymbol{\theta}),
\end{align}
as the maximizer of the MAP function
\begin{align}\label{def:map:function}
f_\mathrm{MAP}(\boldsymbol{\theta})=\frac{|\herm{ \boldsymbol{s}}(\boldsymbol{\theta})\boldsymbol{R}_{\boldsymbol{\eta}}^{-1}\boldsymbol{y}|^2}{\herm{ \boldsymbol{s}}(\boldsymbol{\theta})\boldsymbol{R}_{\boldsymbol{\eta}}^{-1}\boldsymbol{s}(\boldsymbol{\theta})} - \frac{(\tau-\mu_\tau)^2}{\sigma_\tau^2} -\frac{(\nu-\mu_\nu)^2}{\sigma_\nu^2}.
\end{align}
\subsubsection{MAP Ambiguity Function}
On the basis of \eqref{def:map:function}, we define a function 
\begin{equation}\label{def:map:ambiguity:function}
\psi_\mathrm{MAP}(\boldsymbol{\theta}) = f_\mathrm{MAP}(\boldsymbol{\theta})|_{\boldsymbol{y} = \boldsymbol{s}(\boldsymbol{0})},
\end{equation}
closely related to the ambiguity function \eqref{def:ambiguity:function}, which we refer to as the MAP ambiguity function (MAP-AF). This measure is derived directly from the likelihood function by an estimation theoretic argument and has the advantage of incorporating the prior knowledge about the unknown parameters $\boldsymbol{\theta}$ into the ambiguity function. Further, in contrast to the standard ambiguity function \eqref{def:ambiguity:function}, the uncertainty about the attenuation $\gamma$ as well as the effect of colored noise with covariance $\boldsymbol{R}_{\boldsymbol{\eta}}$  is considered by the suggested ambiguity characterization \eqref{def:map:ambiguity:function}.
\begin{figure}[!htbp]
	\centering
	\subfloat[Optimized ($L=1$, $\alpha=\alpha^*$) \label{fig:Opt_MAP_Func}]{\includegraphics{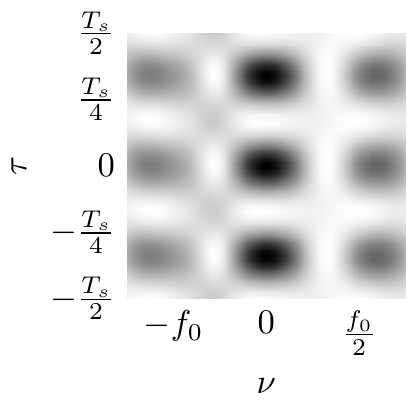}}
	\subfloat[Rectangular \label{fig:Rect_MAP_Func}]{\includegraphics{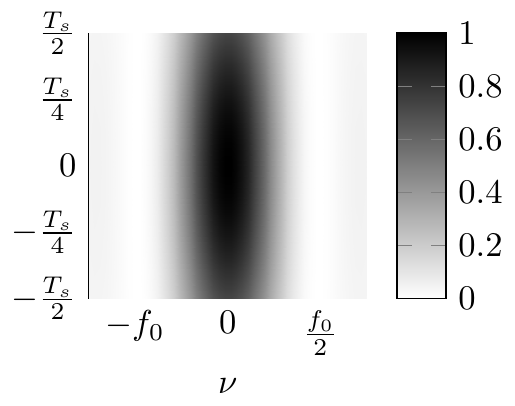}}
	\caption{MAP-AF for different transmit/receive filters}
\end{figure}

Figs. \ref{fig:Opt_MAP_Func} and \ref{fig:Rect_MAP_Func} show the normalized MAP-AF for the optimized transmit and receive filter and for the reference rectangular pulse \eqref{def:reference:rpc} in a signal-dominant environment, i.e., $N_0 \rightarrow 0$ to highlight the effect of the transmit and receive filters. It can be seen that the optimized system exhibits a MAP-AF maximum that is significantly narrower than that of the LFM signal which allows for a higher time-delay and Doppler estimation accuracy. Note that the optimized system shows local maxima, which can cause ambiguities for the solution of the estimator \eqref{def:MAP:Doppler:delay}. However, for a priori knowledge with small uncertainties, i.e., $\sigma_\nu \ll f_0$, $\sigma_\tau \ll T_s$, these local extrema do not influence the estimation as the probability that a MAP-AF function side-lobe exceeds the main lobe for a particular noise realization vanishes.
\section{Simulation Results}
\label{sec:simulation:results}
\begin{figure}[h]
	\centering
	\includegraphics{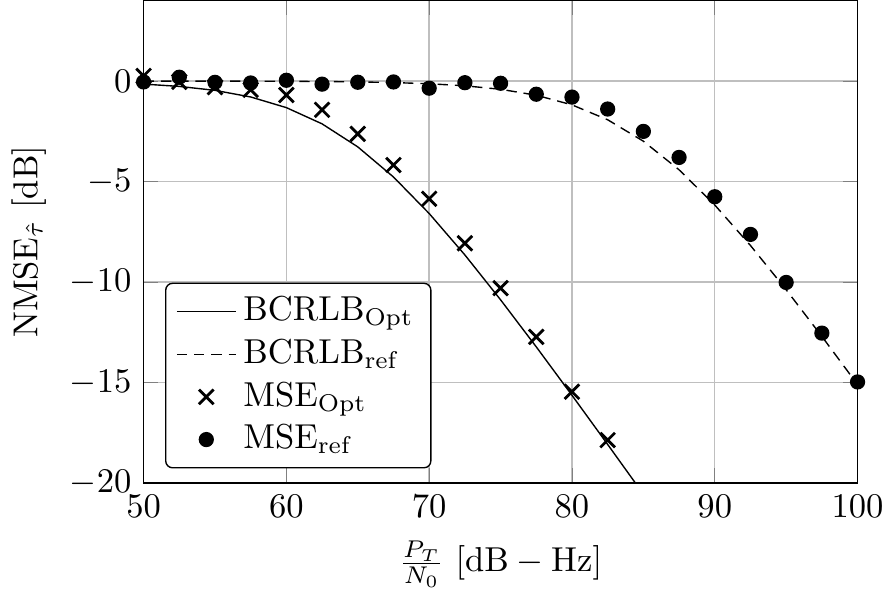}
	\caption{MSE and BCRLB - Time-delay $\tau$}
	\label{fig:MSE_Tau_Equal_Weight}
\end{figure}

\begin{figure}[h]
	\centering
	\includegraphics{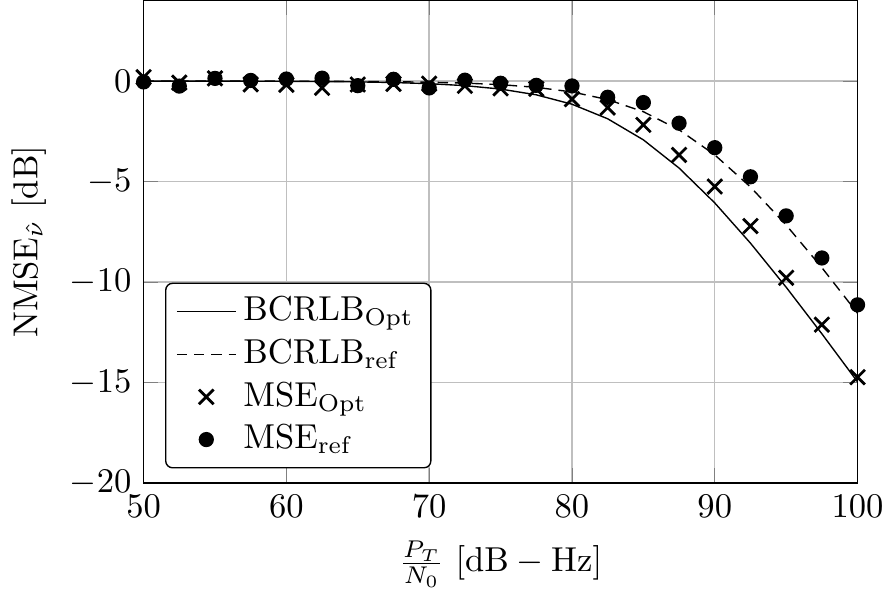}
	\caption{MSE and BCRLB - Doppler shift $\nu$}
	\label{fig:MSE_Nu_Equal_Weight}
\end{figure}

With the estimator \eqref{def:MAP:Doppler:delay} it is possible to use Monte-Carlo simulations to verify the achievable performance gain with an optimized transceiver design. To this end, we use $f_s = 25 \mathrm{MHz}$ and $T_0 = 2\mu \mathrm{s}$. The uncertainty of the time and Doppler shift are $\sigma_\tau= 1 \mathrm{ns}$ and $\sigma_\nu = 5 \mathrm{kHz}$ and $L=1$. The optimized transmit and receive filters are obtained by using the max-sum weighting $\alpha^{\star}$ for the matrix $\ve{M'}$. Figs. \ref{fig:MSE_Tau_Equal_Weight} and \ref{fig:MSE_Nu_Equal_Weight} show the empirical normalized mean squared error (NMSE)
\begin{equation}
\mathrm{NMSE}_{\hat{\tau}/\hat{\nu}} = \frac{\mathrm{MSE}_{\hat{\tau}/\hat{\nu}}}{\sigma^2_{\tau/\nu}}
\end{equation}
of the estimator \eqref{def:MAP:Doppler:delay} for the optimized and rectangular reference systems compared with the analytical BCRLB \eqref{eq:BCRLB} on a double logarithmic scale, where $\mathrm{MSE}_{\hat{\tau}/\hat{\nu}}$ represents the diagonal elements of \eqref{eq:MSE:matrix}, empirically evaluated based on the estimation in \eqref{def:jmap:ml}.
\setlength{\figurewidth}{0.40\textwidth}
\setlength{\figureheight}{4.9cm}
Note that we use the per-Hz signal-to-noise ratio (pSNR) $\frac{P_T}{N_0}$, which is independent of the receive filter, to establish a fair comparison between different transceivers. For low pSNR, the BCRLB saturates since the receive data is so noisy that it does not contribute to the performance of the estimator \eqref{def:MAP:Doppler:delay}. The estimator then exclusively operates on the basis of the prior information and simply produces the mean $\boldsymbol{\hat{\theta}} = \boldsymbol{\mu}_{\boldsymbol{\theta}}$. In the high pSNR regime, it can be observed that the RMSE of the estimator \eqref{def:MAP:Doppler:delay} converges to the BCRLB. In this region, the presented transmit and receive filter optimization is beneficial, as  the estimator favors the received data in the estimation process. Furthermore, it is observed that for moderate to high pSNR values the performance gain is roughly $20\mathrm{dB}$ for the estimation of the time-delay and $4\mathrm{dB}$ for the Doppler estimation, which corresponds to the findings from the Pareto region in Fig. \ref{fig:Pareto_Different_Bandwidths}. The pSNR value at which the receive data starts to contribute to the reduction of the estimation uncertainty is approximately $\mathrm{pSNR}_\tau = 65\mathrm{dB}-\mathrm{Hz}$ for the time-delay and $\mathrm{pSNR}_\nu = 85\mathrm{dB}-\mathrm{Hz}$ for the Doppler shift. The difference between these two pSNR levels can be explained by the different sensitivity of the transceiver with respect to the two unknown parameters. Note that when using a larger transmit signal duration $T_0$, the estimation accuracy of the time-delay increases linearly. In contrast, the Doppler shift performance grows quadratically with $T_0$, since the derivative of the receive signal with respect to the Doppler shift scales linearly with the observation interval. As a consequence, the $\mathrm{pSNR}_\tau$ and $\mathrm{pSNR}_\nu$ threshold level can be aligned by using a larger sampling period $T_0$. During our simulations, we observed that the system optimization has no influence on the estimation performance of the linear path gain $\gamma$.
\section{Conclusion}
\label{sec:conclusion}
We have presented a framework for the joint optimization of the analog transmit and receive filters for estimating the time and Doppler shift with sub-Nyquist sampling. The initial weighted MSE minimization problem has been reformulated as a maximization problem based on the expected Fisher information matrix. Exploiting various approximations, we have shown that it is possible to find transceiver filters that perform significantly better than conventional designs using a simple iterative optimization algorithm. We have demonstrated that the resulting waveforms exhibit bandwidths that violate the sampling theorem ($B>f_s$), while the obtained filters have a bandpass structure capturing the entire signal power of the transmit waveform. The Pareto-optimal region exhibits a small trade-off between delay and Doppler estimation and is substantially increased by jointly optimizing the transmit and receive filters. Further, we have modified the standard ambiguity function by taking into account noise covariance and prior information on the delay and Doppler. With the resulting MAP ambiguity function, we have illustrated the gain of the proposed optimized system design relative to conventional approaches. To evaluate the performance of the adapted sub-Nyquist transceiver, Monte-Carlo simulations were performed. The simulation results show significant accuracy gains, such that we conclude that transceiver optimization is an important aspect when designing parameter estimation systems operating under sub-Nyquist conditions. 
\appendix
\subsection{Fisher Information Matrix for Delay-Doppler Estimation} \label{sec:Fisher_Information_Delay-Doppler_Channel}
The FIM elements \eqref{eq:FI_Entries_General} are obtained by computing the derivatives $\frac{\partial}{\partial [\boldsymbol{\theta}]_i}\boldsymbol{v}(\boldsymbol{\theta})$. Using \eqref{eq:transmit:signal:sampled:filter:fdomain}, we obtain
\begin{align}
\frac{\partial}{\partial \tau} \boldsymbol{v}(\boldsymbol{\theta}) =& \gamma\sqrt{N} \boldsymbol{W}^{\mathrm{H}} \boldsymbol{\tilde{\Delta}}(\nu) \boldsymbol{A} \boldsymbol{\partial T}(\tau) \big(\boldsymbol{\tilde{g}} \circ \boldsymbol{\tilde{h}}(\nu) \big),
\label{eq:D_V_D_Tau} \\
\frac{\partial}{\partial \nu} \boldsymbol{v}(\boldsymbol{\theta}) =& \gamma \sqrt{N} \boldsymbol{W}^\mathrm{H} \Big(\boldsymbol{\partial\tilde{\Delta}}(\nu) \boldsymbol{A} \boldsymbol{T}(\tau) \big(\boldsymbol{\tilde{g}} \circ \boldsymbol{\tilde{h}}(\nu) \big) + \nonumber \\  & \qquad \qquad \,\, \boldsymbol{\tilde{\Delta}}(\nu) \boldsymbol{A} \boldsymbol{T}(\tau) \big(\boldsymbol{\tilde{g}} \circ \boldsymbol{\partial\tilde{h}}(\nu) \big) \Big), \label{eq:D_V_D_Nu}
\end{align}
\begin{figure*}[b]
	\begin{align}
	[\boldsymbol{J}_F(\boldsymbol{\theta})]_{11}&= 2N |\gamma|^2 \left( \boldsymbol{\tilde{\Delta}}(\nu) \boldsymbol{A}  \boldsymbol{\partial T}(\tau) \boldsymbol{v}(\nu)\right)^\mathrm{H} \boldsymbol{\tilde{R}}_{\boldsymbol{\eta}}^{-1}	\left( \boldsymbol{\tilde{\Delta}}(\nu) \boldsymbol{A}\boldsymbol{\partial T}(\tau) \boldsymbol{v}(\nu)\right) \label{eq:FIM_11} \\
	[\boldsymbol{J}_F(\boldsymbol{\theta})]_{12}&=2N |\gamma|^2 \mathrm{Re} \bigg\{ \left( \boldsymbol{\tilde{\Delta}}(\nu)  \boldsymbol{A}  \boldsymbol{\partial T}(\tau) \boldsymbol{v}(\nu)\right)^\mathrm{H} \boldsymbol{\tilde{R}}_{\boldsymbol{\eta}}^{-1} \Big(\boldsymbol{\partial\tilde{\Delta}}(\nu) \boldsymbol{A} \boldsymbol{T}(\tau) \boldsymbol{v}(\nu) + \boldsymbol{\tilde{\Delta}}(\nu) \boldsymbol{A} \boldsymbol{T}(\tau) \boldsymbol{\partial v}(\nu)\Big) \bigg\}  =[\boldsymbol{J}_F(\boldsymbol{\theta})]_{21}\label{eq:FIM_12} \\
	[\boldsymbol{J}_F(\boldsymbol{\theta})]_{22} &= 2 N |\gamma|^2 \Big(\boldsymbol{\partial\tilde{\Delta}}(\nu) \boldsymbol{A} \boldsymbol{T}(\tau) \boldsymbol{v}(\nu)+ \boldsymbol{\tilde{\Delta}}(\nu) \boldsymbol{A} \boldsymbol{T}(\tau) \boldsymbol{\partial v}(\nu)\Big)^\mathrm{H}\boldsymbol{\tilde{R}}_{\boldsymbol{\eta}}^{-1} \Big(\boldsymbol{\partial \tilde{\Delta}}(\nu) \boldsymbol{A} \boldsymbol{T}(\tau) \boldsymbol{v}(\nu) + \boldsymbol{\tilde{\Delta}}(\nu) \boldsymbol{A} \boldsymbol{T}(\tau) \boldsymbol{\partial v}( \nu)\Big) \label{eq:FIM_22}
	\end{align}
\end{figure*}
with the partial derivatives
$\boldsymbol{\partial T}(\tau) = \frac{\partial}{\partial \tau} \boldsymbol{T}(\tau)$, $\boldsymbol{\partial\tilde{\Delta}}(\nu) = \frac{\partial}{\partial \nu} \boldsymbol{\tilde{\Delta}}(\nu)$ and $\boldsymbol{\partial \tilde{h}}(\nu) = \frac{\partial}{\partial \nu} \boldsymbol{\tilde{h}}(\nu)$. Inserting \eqref{eq:D_V_D_Tau} and \eqref{eq:D_V_D_Nu} in \eqref{eq:FI_Entries_General} yields the FIM entries in \eqref{eq:FIM_11}--\eqref{eq:FIM_22}.

\section*{Acknowledgment}
The authors would like to thank the anonymous reviewers for their valuable comments and suggestions to improve the quality of the paper. They are also grateful to Dr.-Ing. Amine Mezghani for many helpful discussions.
\bibliography{IEEEabrv,References.bib}
\bibliographystyle{IEEEtran}
\end{document}